\documentclass[reprint,aps,prl,floatfix,twocolumn,superscriptaddress,showpacs,amsmath,amssymb,showpacs,]{revtex4-1}
\usepackage[utf8]{inputenc}
\usepackage{amsfonts,amsmath,amssymb,amsthm} 
\usepackage{wasysym} 
\usepackage{bbold}
\usepackage[usenames,dvipsnames]{xcolor}
\usepackage[pdftex]{graphicx}
\usepackage{bm}
\usepackage[pdftex]{hyperref}
{\end{pmatrix}\end{medsize}} 
\usepackage{nccmath} 
\usepackage{natbib} 
\usepackage{physics} 

\hypersetup{colorlinks=true,linkcolor=red, citecolor=blue, urlcolor=blue,colorlinks=true,bookmarks=true,breaklinks=true}

\usepackage{array}
\newcolumntype{L}[1]{>{\raggedright\let\newline\\\arraybackslash\hspace{0pt}}m{#1}}
\newcolumntype{C}[1]{>{\centering\let\newline\\\arraybackslash\hspace{0pt}}m{#1}}
\newcolumntype{R}[1]{>{\raggedleft\let\newline\\\arraybackslash\hspace{0pt}}m{#1}}

\begin{document}
\title{Nonequilibrium Two-Particle Self-Consistent Approach}
\author{Olivier \surname{Simard}}
\affiliation{Department of Physics, University of Fribourg, 1700 Fribourg, Switzerland}
\author{Philipp \surname{Werner}}
\affiliation{Department of Physics, University of Fribourg, 1700 Fribourg, Switzerland}
\date{\today}

\begin{abstract}
We present the nonequilibrium implementation of the two-particle self-consistent (TPSC) approach, which has been shown to provide a reliable equilibrium description of interacting lattice systems in the weak- and intermediate-correlation regime. This method captures the effects of local and nonlocal correlations in two- and higher-dimensional systems and satisfies the Mermin-Wagner theorem. We demonstrate the versatility of nonequilibrium TPSC with calculations of the time-dependent spin and charge response functions and the evolution of effective temperatures extracted from different correlation functions, after interaction ramps in the two-dimensional Hubbard model. 
\end{abstract}
\maketitle


{\it Introduction --}
Nonlocal correlations play an important role in low-dimensional lattice systems and in the vicinity of phase transitions or crossovers~\cite{Rohringer2018,Bergeron_2011_optical_cond,kauch_pitons_2019,PhysRevB.103.104415_Simard_pi_ton,kusunose_influence_2006,RevModPhys.77.1027}, but calculating the single- and two-particle correlations in a consistent manner is a challenging task. While strong local or short-ranged correlations can be captured by Dynamical Mean Field Theory (DMFT) \cite{Georges_1996} and its cluster extensions \cite{RevModPhys.77.1027}, the study of long-ranged correlations requires diagrammatic extensions such as the Dynamical Vertex Approximation (D$\Gamma$A) ~\citep{PhysRevB.75.045118_dynamical_vertex_approx} or Dual Boson (DB) ~\citep{RUBTSOV20121320_dual_boson} methods. In the weak-to-intermediate correlation regime, the GW method \cite{Hedin1965}, T-matrix approach \cite{Fukuyama1990}, or Fluctuation Exchange Approximation (FLEX)~\citep{BICKERS1989206_FLEX} are often used, but their suitability for model calculations has been questioned \cite{Gukelberger2015} in a systematic comparison to diagrammatic Monte Carlo results  \cite{VanHoucke2010}. Much more convincing results have recently been demonstrated in benchmarks \cite{PhysRevX.11.011058} of the Two-Particle Self-Consistent (TPSC) approach~\citep{PhysRevB.49.13267_tpsc_1994,tpsc_1997}, which ensures consistency between one-particle and two-particle quantities and satisfies the Mermin-Wagner theorem. This makes TPSC particularly appealing for the study of two-dimensional lattice systems. TPSC captures the pseudo-gap physics~\cite{PhysRevLett.93.147004} and the growth of antiferromagnetic correlations in the renormalized classical regime, where the antiferromagnetic correlation length becomes larger than the de~Broglie wave length. It can also be extended to treat symmetry-broken states~\citep{PhysRevB.68.174502_tpsc_superconductivity} and two-particle vertex corrections~\citep{Bergeron_2011_optical_cond}. 

Describing the interplay between different types of correlations in out-of-equilibrium states adds another layer of complexity. Several of the above-mentioned equilibrium techniques have over the last years been adapted to nonequilibrium setups~\citep{PhysRevB.83.195132_CPT_Potthoff,DB_noneq_2011,PhysRevLett.118.246402_GW_Noneq,PhysRevB.100.235117,PhysRevB.104.245127_non_eq_piton_simard,Stahl2021,Bittner2021,PhysRevLett.128.016801,Hofmann_2016,PhysRevB.90.075117_noneq_DCA,Eckstein2016,PhysRevB.101.085127_bittner_DCA,PhysRevB.102.235169,PhysRevB.89.075306,PhysRevA.92.033419}, but the development of accurate, yet computationally efficient nonequilibrium methods for two- and three-dimensional lattice systems remains a challenging research frontier. Here, we introduce the nonequilibrium TPSC method as a powerful addition to the existing toolset, especially in the moderate-correlation regime. We describe the implementation on the Kadanoff-Baym contour and demonstrate the usefulness of the scheme with interaction-quenches in the two-dimensional Hubbard model.


{\it Model and methods --}
We consider the single-band Hubbard model on a $D$-dimensional lattice 
\begin{align}
\label{eq:Hubbard_model_intro}
\hat{\mathcal{H}}(t)=&-\sum_{i,j,\sigma}t^{\text{hop}}_{i,j}\left(\hat{c}_{i,\sigma}^{\dagger}\hat{c}_{j,\sigma}+\text{H.c.}\right)+U(t)\sum_i\hat{n}_{i,\uparrow}\hat{n}_{i,\downarrow}\nonumber\\
&-\mu \sum_i (\hat{n}_{i,\uparrow} + \hat{n}_{i,\downarrow}),
\end{align} 
where $t_{i,j}^\text{hop}$ denotes the hopping amplitude, and $i,j$ refer to lattice sites. The spin is denoted by $\sigma \in \{\uparrow,\downarrow\}$, and $\hat{c}^{(\dagger)}_{i,\sigma}$ are the annihilation (creation) operators for site $i$. $\hat{n}_{i\sigma}=\hat{c}^{\dagger}_{i,\sigma}\hat{c}_{i,\sigma}$ is the number operator, $U(t)$ is the local Hubbard repulsion and $\mu$ the chemical potential. In the applications, we will show results for a two-dimensional lattice with nearest-neighbor hopping $t_{\text{hop}}$ and use $t_\text{hop}$ as the unit of energy ($\hbar/t_\text{hop}$ as the unit of time). We set $\hbar$, $k_B$, the electric charge $e$ and the lattice spacing equal to unity. 

The single- and two-particle correlation functions in TPSC are derived from the following ansatz for the Luttinger-Ward functional $\Phi$ \citep{PhysRevB.49.13267_tpsc_1994,tpsc_1997},

\begin{align}
\label{eq:luttinger_ward_functional}
\Phi[\mathcal{G}] &= \frac{1}{2}\int_{\mathcal{C}}\mathrm{d}z \ \sum_{\sigma}\mathcal{G}_{\sigma}(z,z^+)\Gamma_{\sigma,\sigma}(z)\mathcal{G}_{\sigma}(z,z^+)\notag\\
&+\frac{1}{2}\int_{\mathcal{C}}\mathrm{d}z \ \sum_{\sigma}\mathcal{G}_{\sigma}(z,z^+)\Gamma_{\sigma,-\sigma}(z)\mathcal{G}_{-\sigma}(z,z^+),
\end{align}
where $\mathcal{C}$ represents the Kadanoff-Baym (KB) contour~\citep{kadanoff_baym_1962_noneq}, $z\in \mathcal{C}$ and $z^+$ is infinitesimally \textit{later} than $z$ in the contour ordering. $\mathcal{G}$ represents the one-body Green's function and $\Gamma$ the local irreducible vertices in the particle-hole channel (longitudinal and transversal). The self-energy functional $\Sigma[\mathcal{G}]$ can be derived from $\Phi[\mathcal{G}]$ as $\Sigma_{ij}(z,z^{\prime}) = \frac{\delta\Phi[\mathcal{G}]}{\delta\mathcal{G}_{ji}(z^{\prime},z)}$. The irreducible vertex in the spin channel $\Gamma^{\text{sp}}$ is defined as $\Gamma^{\text{sp}}\equiv \Gamma_{\sigma,-\sigma}-\Gamma_{\sigma,\sigma}$, 
with $\Gamma_{\sigma,-\sigma}=-\frac{\delta\Sigma_{\sigma}}{\delta\mathcal{G}_{-\sigma}}$ and $\Gamma_{\sigma,\sigma}=-\frac{\delta\Sigma_{\sigma}}{\delta\mathcal{G}_{\sigma}}$,
while the irreducible vertex in the charge channel $\Gamma^{\text{ch}}$ is defined as $\Gamma^{\text{ch}}\equiv \Gamma_{\sigma,-\sigma}+\Gamma_{\sigma,\sigma}$.

The ansatz \eqref{eq:luttinger_ward_functional} and the equations of motion for Eq.~\eqref{eq:Hubbard_model_intro} lead to the following approximate relation between the two-particle density-density correlation function (double occupancy) and the single-particle self-energy and Green's function, 

\begin{align}
\label{eq:first_approx_TPSC_two_particle}
&\Sigma_{l\bar{b};\sigma\bar{\sigma}^{\prime}}(z_1,\bar{z})\mathcal{G}_{\bar{b}m;\bar{\sigma}^{\prime}\sigma}(\bar{z},z_2)\notag\\
&\simeq -iU(z_1)\frac{\left<\hat{n}_{l,-\sigma}(z_1)\hat{n}_{l,\sigma}(z_1)\right>}{\left<\hat{n}_{l,-\sigma}(z_1)\right>\left<\hat{n}_{l,\sigma}(z_1)\right>}\biggl(\mathcal{G}_{ll;-\sigma}(z_1,z_1^+)\notag\\
&\hspace{4mm}\times\!\mathcal{G}_{lm;\sigma}(z_1,z_2)\!-\!\mathcal{G}_{ll;\sigma-\sigma}(z_1,z_1^+)\mathcal{G}_{lm;-\sigma\sigma}(z_1,z_2)\!\biggr),
\end{align}
where, from now on, identical contour-ordered variables and indices featuring an overline are integrated/summed over, and Latin subscripts represent lattice sites. The angular brackets denote the grand-canonical ensemble average $\left<\cdots\right>\equiv\frac{1}{\mathcal{Z}}\sum_{i}\bra{\Psi_i}\mathcal{T}_{\mathcal{C}}e^{-i\int_{\mathcal{C}}\mathrm{d}\bar{z}\hat{\mathcal{H}}(\bar{z})} \cdots\ket{\Psi_i}$, with $\mathcal{T}_{\mathcal{C}}$ the time-ordering operator on $\mathcal{C}$, $\{\ket{\Psi_i}\}$ a set of states spanning the Fock space and $\mathcal{Z}=\sum_{i}\bra{\Psi_i}e^{-i\int_{\mathcal{C}}\mathrm{d}\bar{z}\hat{\mathcal{H}}(\bar{z})}\ket{\Psi_i}$ the partition function. The interaction $U(z)$ is a function on the contour $\mathcal{C}$ that is related to $U(t)$ in Eq.~\eqref{eq:Hubbard_model_intro} as follows: on the imaginary-time branch $U(z)=U(t=0^-)$ is the interaction in the initial equilibrium state, whereas on the real-time branches $U(z)=U(t) \ \forall \ t\geq 0$. Equation~\eqref{eq:first_approx_TPSC_two_particle} becomes exact when the substitutions $z_2\to z_1^+$ and $m\to l$ are made.

Using the definition of $\Gamma^{\text{sp}}$ it follows from Eq.~\eqref{eq:first_approx_TPSC_two_particle} that 
\begin{align}
\label{eq:gamma_sp_vertex}
i\Gamma^{\text{sp}}(z)=U(z)\frac{\left<\hat{n}_{-\sigma}(z)\hat{n}_{\sigma}(z)\right>}{\left<\hat{n}_{-\sigma}(z)\right>\left<\hat{n}_{\sigma}(z)\right>}.
\end{align}
The irreducible vertices in both the spin and charge channels obey the fluctuation-dissipation theorem for two-particle correlation functions
\begin{align}
\label{eq:fluctuation_dissipation_two_particle}
&i\int_{-\pi}^{\pi}\frac{\mathrm{d}^Dq}{\left(2\pi\right)^D} \ \chi^{\text{sp/ch}}_{\mathbf{q}}(z,z^{+})\notag\\
&\quad =n(z)+2(-1)^l\left<\hat{n}_{-\sigma}(z)\hat{n}_{\sigma}(z)\right> - (1-l)n(z)^2,
\end{align}
where $n=\left<\hat{n}_{\uparrow}+\hat{n}_{\downarrow}\right>$ is the density of particles, $l=0$ for charge (ch) and $l=1$ for spin (sp). The spin (charge) susceptibility is denoted by $\chi^{\text{sp}(\text{ch})}$. The susceptibilities obey the Bethe-Salpether equation
\begin{align}
\label{eq:bethe_salpether_eq}
&\chi_{\mathbf{q}}^{\text{sp/ch}}(z,z^{\prime}) = \chi^0_{\mathbf{q}}(z,z^{\prime})\notag\\ 
&\quad + (-1)^{l+1}\frac{i}{2} \chi_{\mathbf{q}}^0(z,\bar{z})\Gamma^{\text{sp}/\text{ch}}(\bar{z})\chi_{\mathbf{q}}^{\text{sp/ch}}(\bar{z},z^{\prime}),
\end{align}
where $\chi^0$ is the non-interacting susceptibility 
\begin{align}
\label{eq:non_interacting_susceptibility}
\chi^0_{\mathbf{q}}(z,z^{\prime}) = -2i\int_{-\pi}^{\pi}\frac{\mathrm{d}^Dk}{(2\pi)^D} \ \mathcal{G}^0_{\mathbf{k}}(z,z^{\prime}) \mathcal{G}^0_{\mathbf{k}+\mathbf{q}}(z^{\prime},z).
\end{align} 
Eqs.~\eqref{eq:gamma_sp_vertex}-\eqref{eq:non_interacting_susceptibility}
self-consistently fix $\Gamma^{\text{sp}}$. Once $\Gamma^{\text{sp}}$ is known, $\Gamma^{\text{ch}}$ can be determined using Eqs.~\eqref{eq:fluctuation_dissipation_two_particle} and \eqref{eq:bethe_salpether_eq}, since knowing $\Gamma^{\text{sp}}$ fixes the double occupancy via Eq.~\eqref{eq:gamma_sp_vertex}.

Using generating functionals with source fields $\phi_{\sigma,\sigma^{\prime}}(z,z^{\prime})$ and the equations of motion for Eq.~\eqref{eq:Hubbard_model_intro}, one can compute the second-level approximation $\Sigma^{(1)}$ to the self-energy~\citep{tpsc_1997}. By approximating the vertices composing $\Sigma^{(1)}$ as fully local, $\Gamma_{l,m,i,j}(z_1,z_2;z_3,z_4) \sim \Gamma_m(z_2)\delta_{\mathcal{C}}(z_2,z_1)\delta_{\mathcal{C}}(z_2,z_3)\delta_{\mathcal{C}}(z_2^+,z_4)\delta_{l,m}\delta_{l,i}\delta_{l,j}$ \footnote{$\delta_{\mathcal{C}}(z,z^{\prime})$ are Dirac delta functions on the contour $\mathcal{C}$ and $\delta_{r,s}$ are Kronecker deltas}, $\Sigma^{(1)}$ becomes~\citep{senechal_bourbonnais_tremblay_2004}
\begin{align}
\label{eq:tpsc_self_energy}
&\Sigma^{(1)}_{\mathbf{k}}[\alpha](z,z^{\prime})
= U(z)n(z)\delta_{\mathcal{C}}(z,z^{\prime}) + \frac{U(z)}{8}\int_{-\pi}^{\pi}\frac{\mathrm{d}^Dq}{(2\pi)^D}\alpha(z^{\prime})\notag\\
&\!\!\times\!\!\biggl[3\Gamma^{\text{sp}}(z^{\prime})\chi^{\text{sp}}_{\mathbf{q}}(z^{\prime},z) + \Gamma^{\text{ch}}(z^{\prime})\chi^{\text{ch}}_{\mathbf{q}}(z^{\prime},z)\biggr]\mathcal{G}^0_{\mathbf{k}+\mathbf{q}}(z,z^{\prime}),\hspace{0.1mm}
\end{align} where the one-time variable $\alpha$ has been introduced to satisfy the sum-rule involving the double occupancy
appearing in Eq.~\eqref{eq:first_approx_TPSC_two_particle},
\begin{align}
\label{eq:double_occupancy_sum_rule}
&\frac{-i}{2}\int_{-\pi}^{\pi}\frac{\mathrm{d}^Dk}{(2\pi)^D} \ \left[\Sigma_{\mathbf{k},\bar{\sigma}}^{(1)}[\alpha](z,\bar{z})\mathcal{G}_{\mathbf{k},\bar{\sigma}}[\Sigma^{(1)}](\bar{z},z^{+})\right]\notag\\
& \quad = U(z)\langle \hat n_{-\sigma}(z)\hat n_{\sigma}(z) \rangle.
\end{align}
Equation~\eqref{eq:tpsc_self_energy} preserves the crossing symmetry, that is the symmetry under the exchange of two particles or two holes. Hence, Eq.~\eqref{eq:tpsc_self_energy} is a symmetrized version of the self-energies computed in the longitudinal ($\phi_{\sigma,\sigma}$) and transverse ($\phi_{\sigma,-\sigma}$) particle-hole channels.

We note that Eq.~\eqref{eq:fluctuation_dissipation_two_particle} which fixes $\Gamma^{\text{sp}/\text{ch}}$ is time-local, while Eq.~\eqref{eq:double_occupancy_sum_rule} which determines $\alpha$ involves a convolution on the contour. This results in a qualitatively different time evolution of the two quantities. It is also important to mention that reinserting $\mathcal{G}[\Sigma^{(1)}]$ into Eq.~\eqref{eq:non_interacting_susceptibility} and iterating until convergence improves the energy conservation on the real-time axis, but makes the TPSC procedure violate sum-rules, even though this violation is small~\citep{Vilk_1996,PhysRevB.49.13267_tpsc_1994,tpsc_1997,PhysRevX.11.011058}. The latter self-consistent method has been coined TPSC+GG~\citep{PhysRevX.11.011058}. The sum-rule \eqref{eq:double_occupancy_sum_rule} is enforced for TPSC, but not for TPSC+GG, for numerical stability reasons.

We now summarize the steps in the calculation of the self-energy \eqref{eq:tpsc_self_energy}, which are understood to apply to all 
components of the two-time functions on the KB contour. (i) Compute the noninteracting susceptibility \eqref{eq:non_interacting_susceptibility} from a noninteracting Green's function $\mathcal{G}^0$ whose chemical potential matches the desired electronic density $n$. Note that the chemical potentials for $\mathcal{G}^0$ ($\mu_0$) and $\mathcal{G}$ ($\mu$) are different, but the difference $\mu-\mu_0$ is compensated by the change of the real part of the retarded self-energy at the Fermi surface $\Re\Sigma^{\text{R}}(\mathbf{k}_F,\omega=\mu)$~\citep{tpsc_1997}. (ii) Simultaneously solve the Bethe-Salpether equation \eqref{eq:bethe_salpether_eq} for the spin channel ($l=1$), the fluctuation-dissipation relation \eqref{eq:fluctuation_dissipation_two_particle} for the spin channel, and Eq.~\eqref{eq:gamma_sp_vertex}. Those three equations translate into a root-finding problem: within a time-stepping scheme, we solve for $\Gamma^{\text{sp}}$ at each new maximum contour-time $z$ using a multidimensional Newton-Raphson method for nonlinear systems of equations. This determines $\Gamma^{\text{sp}}(z)$, $\chi^{\text{sp}}(z,z^{\prime})$ and the double occupancy $\langle \hat{n}_{-\sigma}(z)\hat{n}_{\sigma}(z)\rangle$. (iii) With the double occupancy known, the Bethe-Salpether equation \eqref{eq:bethe_salpether_eq} for the charge channel ($l=0$) and Eq.~\eqref{eq:fluctuation_dissipation_two_particle} for the charge channel are simultaneously solved to get $\Gamma^{\text{ch}}(z)$ and $\chi^{\text{ch}}(z,z^{\prime})$, still using the time-stepping scheme and a multidimensional root-finding method. (iv) Compute $\Sigma^{(1)}(z,z^{\prime})$ using Eq.~\eqref{eq:tpsc_self_energy} (in conjunction with the sum-rule \eqref{eq:double_occupancy_sum_rule} in TPSC). TPSC+GG contains an extra loop where $\mathcal{G}[\Sigma^{(1)}]$ is reinserted into Eq.~\eqref{eq:non_interacting_susceptibility} and the full procedure described above is repeated until convergence. Our implementation is based on the NESSi library~\citep{Nessi} and uses the irreducible Brillouin zone for integrals in reciprocal space~\citep{2021_irr_BZ_algo}.

\begin{figure}[t]
    \includegraphics[width=\linewidth]{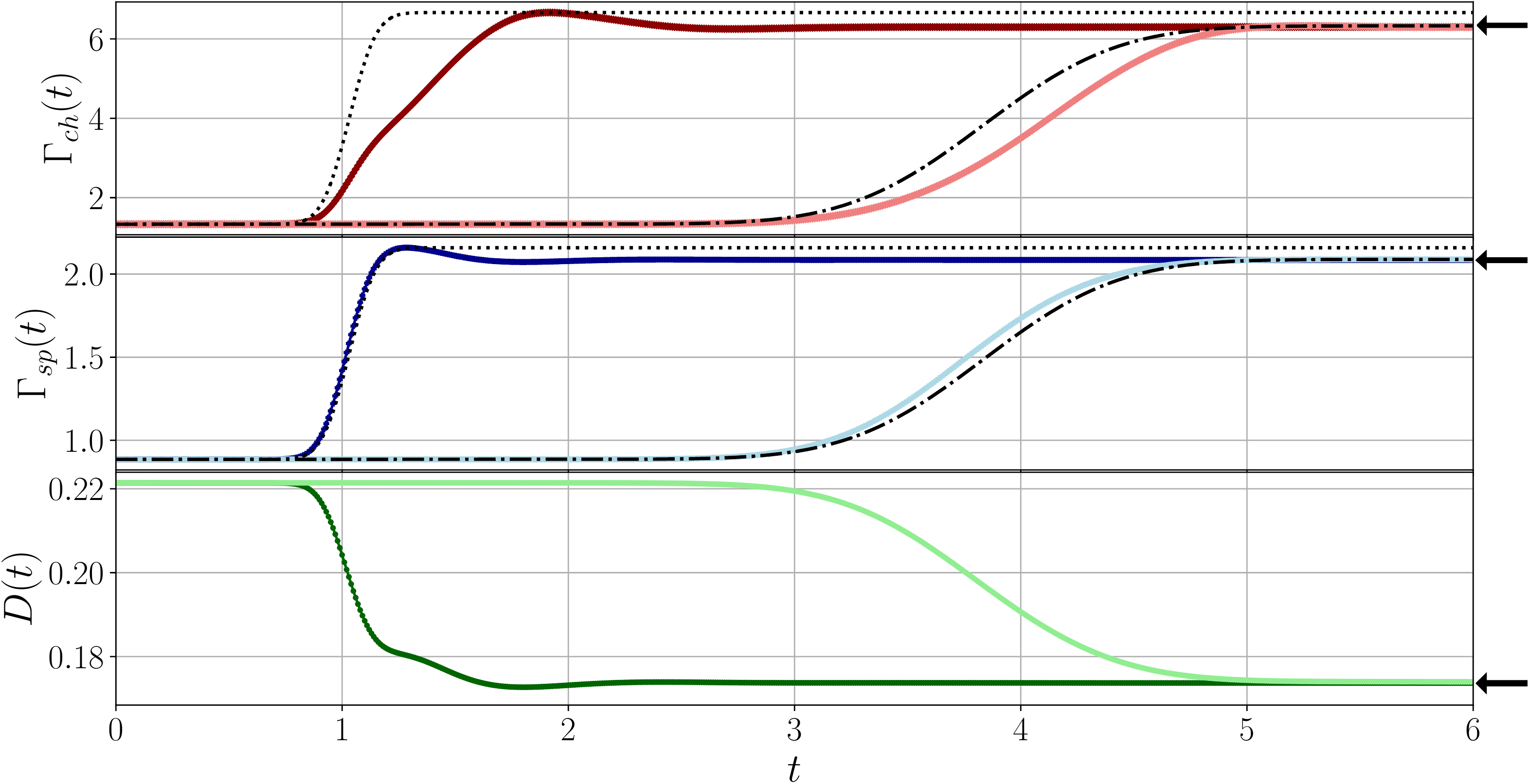}
\caption{Dynamics after interaction ramps from $U=1$ to $U=3$. The dotted and dash-dotted lines show the ramp profiles in arbitrary units. The quantities plotted in light (dark) colors correspond to the slow (fast) ramp. Top panel: charge irreducible vertex $\Gamma^\text{ch}(t)$. Middle panel: spin irreducible vertex $\Gamma^\text{sp}(t)$. Bottom panel: Double occupancy $D(t)$. 
The arrows on the right indicate the thermalized values for the slow ramp ($\beta_\text{th}=2.4$), calculated from $E_\text{tot}$ after the ramp. 
} 
\label{fig:up_ramp_effect_profile}
\end{figure}

{\it Results --}
We apply nonequilibrium TPSC+GG to the half-filled 2D Hubbard model \eqref{eq:Hubbard_model_intro} to study the dynamics induced by interaction ramps. The corresponding TPSC results can be found in the Supplemental Material (SM).
As a first illustration, we show in Fig.~\ref{fig:up_ramp_effect_profile} the evolution of the spin ($\Gamma^\text{sp}(t)$) and charge ($\Gamma^\text{ch}(t)$) vertices, as well as the double occupation $D(t)=\langle n_\uparrow(t)n_\downarrow(t)\rangle$, after ramps from $U=1$ to $U=3$ for the initial inverse temperature $\beta=1/T=3$. Results are reported both for slow and fast ramps, whose profiles are indicated by the black dot-dashed and dotted lines, respectively.

In the case of the fast ramp, one notices an interesting decoupling of the dynamics of $\Gamma^\text{sp}(t)$ and $\Gamma^\text{ch}(t)$. While $\Gamma^\text{sp}(t)$ reacts fast to the change in $U$ -- the spin vertex grows on the timescale set by the ramp -- there is a clear delay in the growth of $\Gamma^\text{ch}$. In the case of $\Gamma^\text{sp}$, the fast response and fast relaxation after the ramp can be explained by Eq.~\eqref{eq:gamma_sp_vertex}: the right-hand side is proportional to $U(t)$, the denominator is constant, and $D(t)$ changes only slightly (by about $10\%$). After the fast ramp, the double occupation continues to decrease up to $t\approx 1.8$, which produces the overshooting of $\Gamma^\text{sp}$. After $t\approx 2$, the double occupation is thermalized and thus also $\Gamma^\text{sp}$ reaches a constant value. On the other hand, for $\Gamma^\text{ch}$, the small change of the double occupation after the ramp still produces (via Eqs.~\eqref{eq:fluctuation_dissipation_two_particle} and \eqref{eq:bethe_salpether_eq}) a substantial increase. The different quantities reach a plateau at large $t$ whose value only slightly depends on the ramp profile: this is because the inverse temperatures $\beta_\text{th}$ of the thermalized systems after the fast and slow ramps do not differ much ($\beta_\text{th}=2.29$ and $\beta_\text{th}=2.4$, respectively). 

Since an interaction ramp injects an energy $\Delta E$ into the system, the temperature of the thermalized state will be higher than that of the initial state. $\beta_\text{th}$ can be computed from the total energy of the system after the ramp. The total energy is given by the sum $E_{\text{tot}}(t)=E_k(t)+E_p(t)$, with the kinetic energy $E_k(t)=\frac{-2i}{N_k}\sum_{\mathbf{k}}\epsilon_{\mathbf{k}}(t)\mathcal{G}^<_{\mathbf{k}}(t,t)$ and the potential energy $E_p(t)=\frac{-i}{N_k}\sum_{\mathbf{k}}\int_{\mathcal{C}}\mathrm{d}z \left[\Sigma_{\mathbf{k}}(t,z)\mathcal{G}_{\mathbf{k}}(z,t)\right]^<$ \footnote{The potential energy can also be obtained from Eq.~\eqref{eq:fluctuation_dissipation_two_particle}. The results are consistent within numerical accuracy.}, where $\epsilon_{\mathbf{k}}$ is the square lattice dispersion and (unless otherwise noted) we evaluate the sums with $N_k=30\times 30$ $\mathbf{k}$-points. Since $E_\text{tot}(t)$ is almost conserved after the ramp, this allows us to compute $\beta_{\text{th}}$ by searching for the $\beta$ of the thermal system with the post-ramp $U$ and energy $E_{\text{tot}}(0)+\Delta E$. The black arrows in Fig.~\ref{fig:up_ramp_effect_profile} show the thermalized values for the slow ramp (those for the fast ramp are almost indistinguishable). The good agreement between the thermal reference data and the long-time values demonstrates that the TPSC dynamics capture the relatively fast thermalization of local quantities after the ramps. 

\begin{figure}
\includegraphics[width=\linewidth]{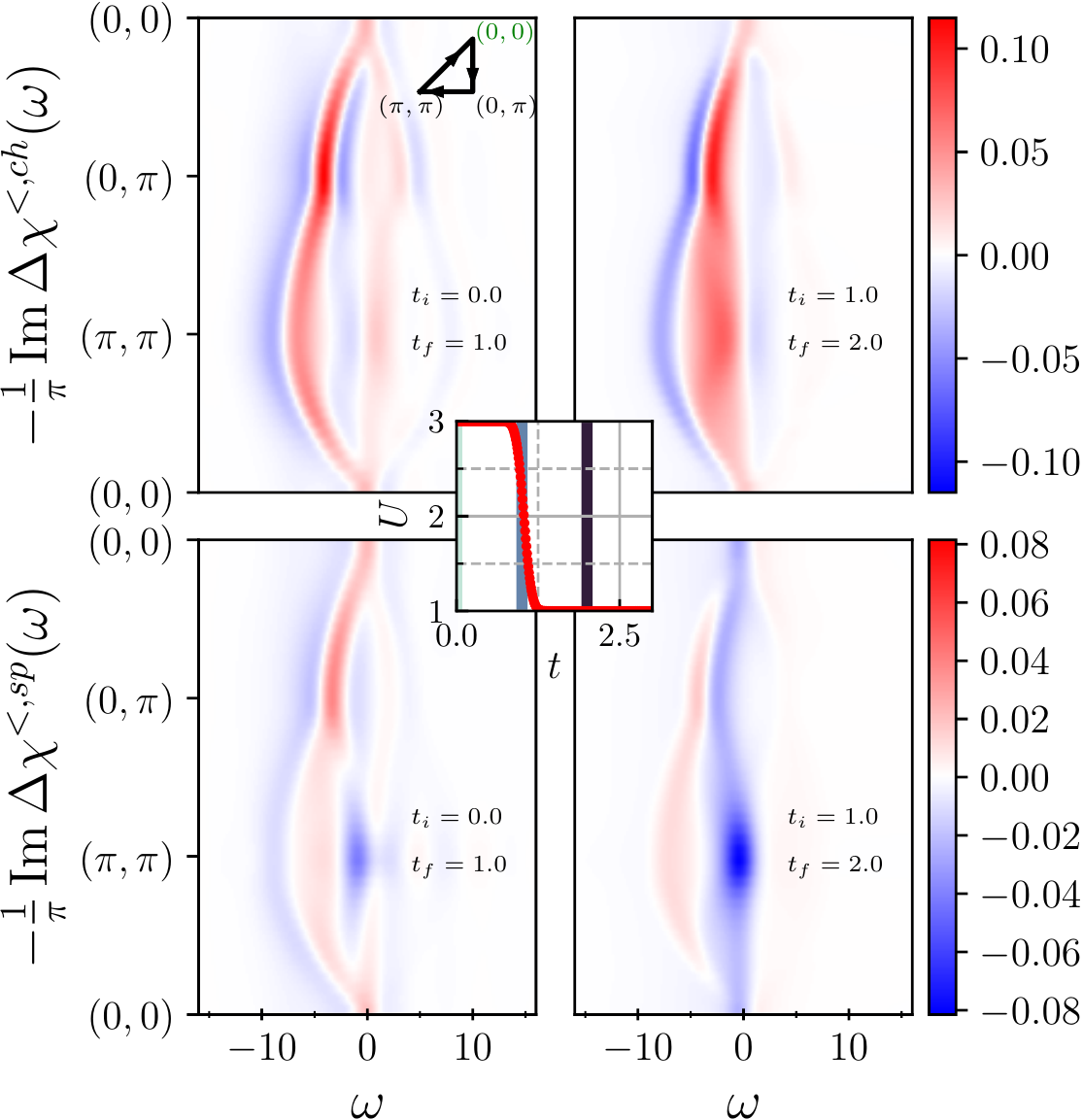}
\caption{ Top (Bottom) panels: Difference spectra of the lesser component of the charge (spin) susceptibility after the interaction ramp shown in the inset. The inset black triangle illustrates the path in reciprocal space along which the spectra are displayed. The times $t_i$ and $t_f$ used in the calculation of the difference spectra are annotated in each panel. The time window used in the Fourier transformation is $\Delta t=4$. Each row of panels uses the same color scale.} 
\label{fig:k_resolved_retarded_charge}
\end{figure}

In Fig.~\ref{fig:k_resolved_retarded_charge} we show the $\mathbf{k}$-resolved time evolution of the spin and charge susceptibilities. Plotted are the imaginary parts of the lesser components of the time differences $\Delta\chi(t_f,t_i,\omega)\equiv\chi(t_i,\omega)-\chi(t_f,\omega)$ for the charge (top panels) and spin (bottom panels) susceptibilities. The initial inverse temperature is $\beta=3$. The profile of the ramp from $U=3$ to $U=1$, with inflection point at $t=1$, is illustrated in the middle inset of Fig.~\ref{fig:k_resolved_retarded_charge}. The left panels show the difference between $t_f=1$ and $t_i=0$, while the right panels show the difference between $t_f=2$ and $t_i=1$ (data for a ramp from $U=1$ to $U=3$ are shown in the SM). The result for $\Delta\chi^\text{ch}$ implies a renormalization of the dispersive features which can be explained by the reduced broadening of the density of states with decreasing $U$. During the ramp, one observes spectral weight near $\omega\simeq 0$ with maximum intensity around $\mathbf{k}=(\pi,\pi)$ and a two-peak structure with inverted weight around $\mathbf{k}=(0,\pi)$, which indicates the transient appearance of low-energy charge excitations. As expected for a ramp to small $U$, the charge susceptibility approaches the result for the simple bubble (Lindhard function)~\cite{PhysRevB.95.165127}.
The bottom panels of Fig.~\ref{fig:k_resolved_retarded_charge} show that the ramp from $U=3$ to $1$ mainly affects the spin excitations around $\mathbf{k}=(\pi,\pi)$, where the spectral weight is strongly reduced. This is expected since the combined effect of the reduced $U$ and the heating suppresses the antiferromagnetic correlations. Both in the charge and spin susceptibility, the larger spectral change is observed in the second half of the ramp. In the SM, we also show the TPSC+GG equilibrium spectra for $U=1$ and $U=3$ at $\beta=3$, the evolution of the single-particle spectra, as well as analogous $\mathbf{k}$-resolved time difference maps for the TPSC scheme.

As a third illustration of the nonequilibrium TPSC method, we study the thermalization of the one- and two-particle quantities $\mathcal{G}$, $\chi^{\text{ch}}$, and $\chi^{\text{sp}}$, and ask to what extent an effective temperature of the nonequilibrium state can be defined. For this purpose, we introduce a frequency and momentum dependent  $\beta_\mathbf{k}(t,\omega)$ measured at time $t$ by the formula~\cite{PhysRevB.102.235169}
\begin{align}
\label{eq:dynamical_inverse_temperature}
\beta_{\mathbf{k}}(t,\omega) = \frac{1}{\omega}\ln\left[\mp\frac{\mathcal{A}^R_{\mathbf{k}}(t,\omega)}{\mathcal{A}^{<}_{\mathbf{k}}(t,\omega)} \pm 1\right],
\end{align}
which in equilibrium reduces to the inverse temperature of the system. In Eq.~\eqref{eq:dynamical_inverse_temperature}, the upper (lower) sign holds for bosonic (fermionic) quantities. If we group $\mathcal{G}$, $\chi^{\text{ch}}$ and $\chi^{\text{sp}}$ under $\Lambda$, then $\mathcal{A}^R_{\mathbf{k}}(t,\omega) = -\frac{1}{\pi}\Im\Lambda^R_{\mathbf{k}}(t,\omega)$ stands for the spectral function, computed with the retarded component, while $\mathcal{A}^{<}_{\mathbf{k}}(t,\omega) = \frac{1}{2\pi}\Im\Lambda^{<}_{\mathbf{k}}(t,\omega)$~\cite{aoki_nonequilibrium_2014}. 

\begin{figure}
\includegraphics[width=\linewidth]{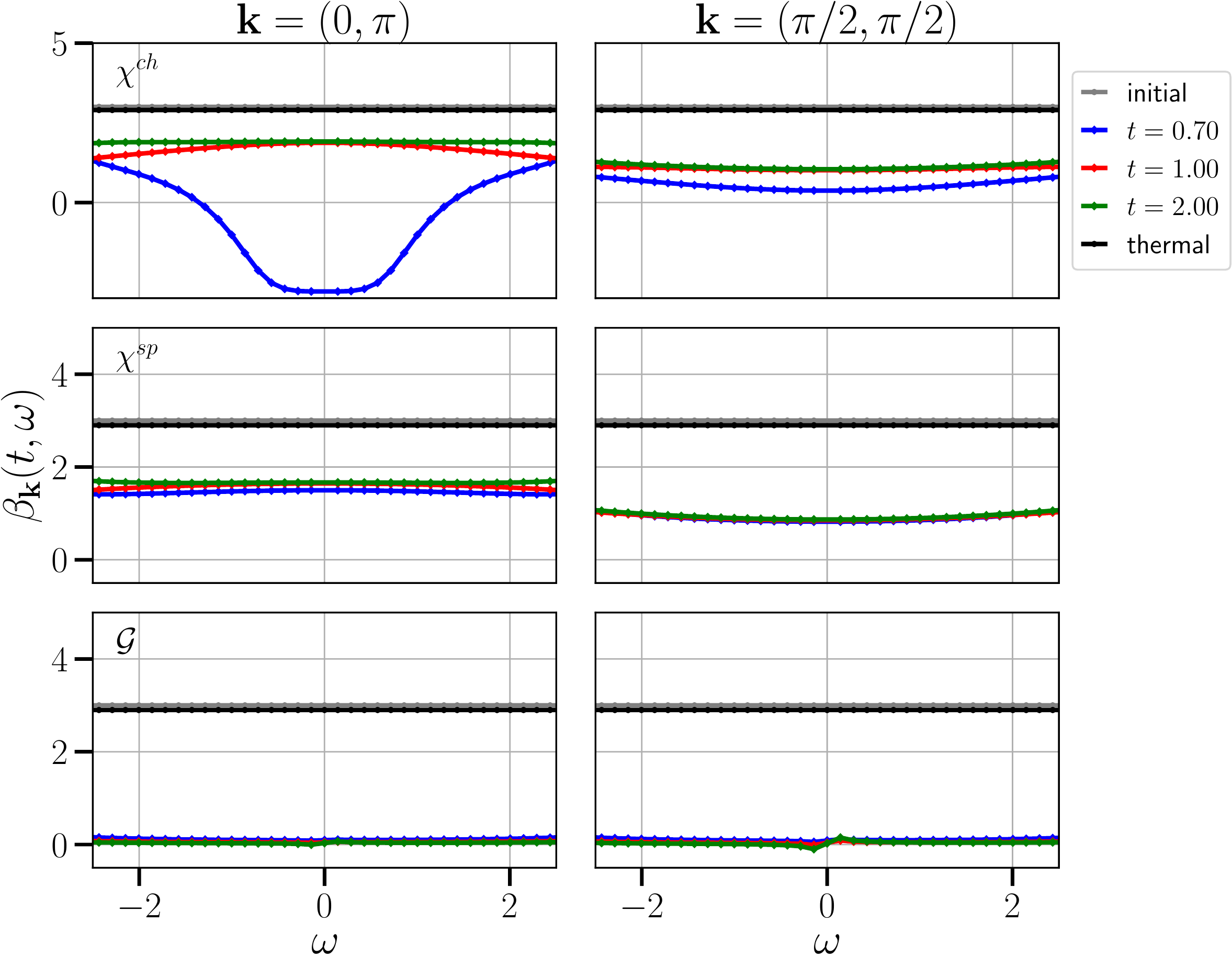}
\caption{Effective inverse temperatures $\beta_\mathbf{k}(t,\omega)$ for the Fermi momenta $\mathbf{k}=(0,\pi)$ (left panels) and $\mathbf{k}=(\frac{\pi}{2},\frac{\pi}{2})$ (see Eq.~\eqref{eq:dynamical_inverse_temperature}). The gray curve plots the initial $\beta=3$, the blue curve $\beta_\mathbf{k}(t=0.7,\omega)$, the red curve $\beta_\mathbf{k}(t=1,\omega)$, the green curve $\beta_\mathbf{k}(t=2,\omega)$, and the black curve the thermalized value $\beta_\text{th}=2.9$. The top panels show the effective inverse temperature extracted from $\chi^{\text{ch}}$, the middle panels for $\chi^{\text{sp}}$ and the bottom panels for $\mathcal{G}$. Here, we use a $40\times 40$ ${\mathbf k}$-grid.
} 
\label{fig:dynamical_inverse_T}
\end{figure}

In Fig.~\ref{fig:dynamical_inverse_T}, we plot $\beta_\mathbf{k}(t,\omega)$ for $\mathbf{k}=(0,\pi)$ (left panels) and $\mathbf{k}=(\frac{\pi}{2},\frac{\pi}{2})$ (right panels) before the interaction ramp ($t=0$), during the ramp ($t=0.7$ and $t=1$), after the ramp ($t=2$) and in the thermalized state. The interaction ramp profile is the one depicted in Fig.~\ref{fig:k_resolved_retarded_charge}. Interestingly, for the two Fermi momenta, one finds that while a nonequilibrium temperature can be defined by Eq.~\eqref{eq:dynamical_inverse_temperature}, in the sense that $\beta_\mathbf{k}(t,\omega)$ varies slowly with $\omega$ near $\omega=0$, all the three quantities exhibit different nonequilibrium temperatures and different relaxations towards the thermal value. For $\mathbf{k}=(0,\pi)$, the charge susceptibility displays a negative temperature in the middle of the interaction ramp ($t=0.7$) around $\omega=0$, which is related to the short-lived transient charge excitations with inverted weight near $\omega\simeq 0$ seen in  Fig.~\ref{fig:k_resolved_retarded_charge}. For $\mathbf{k}=(\frac{\pi}{2},\frac{\pi}{2})$, the charge susceptibility does not yield negative effective $\beta$ for the times considered, although the result for $t=0.7$ corresponds to a high effective temperature and displays a significant $\omega$-dependence. At $t=1$ and later, for both Fermi momenta, the charge susceptibility effective temperature becomes almost $\omega$-independent and slowly approaches the thermal value. The spin susceptibility shows a positive $\beta_\mathbf{k}(t,\omega)$ for all intermediate times ($t=0.7,1,2$) and an even slower relaxation. At $t=2$, the charge and spin susceptibilities yield comparable inverse temperatures for given $\mathbf{k}$, but the results differ between the two momenta. The $\beta_{\mathbf{k}=(\frac{\pi}{2},\frac{\pi}{2})}(t,\omega)$ and  $\beta_{\mathbf{k}=(0,\pi)}(t,\omega)$ extracted from the one-body Green's function correspond to very high effective temperatures, which increase up to $t=2$. The slow thermalization of $\mathcal{G}_\mathbf{k}$ near the Fermi level in weakly correlated systems is expected and has already been discussed in Refs.~\cite{Moeckel_2008,PhysRevB.104.245127_non_eq_piton_simard}.


{\it Conclusions --}
We presented two variants of a promising method for treating nonlocal correlations in two- and higher-dimensional nonequilibrium systems, the nonequilibrium TPSC and TPSC+GG approaches. In equilibrium, these approximate methods yield remarkably accurate results in the intermediate-correlation regime  \cite{PhysRevX.11.011058} and they satisfy the Mermin-Wagner theorem \cite{tpsc_1997}. We explained the implementation on the Kadanoff-Baym contour and applied these methods to interaction ramps in the 2D Hubbard model. Fast perturbations induce qualitatively different dynamics in the spin and charge channels, but the local vertices and double occupation thermalize within a few inverse hopping times. 
Transient low-energy charge excitations corresponding to a negative effective temperature in the charge sector can appear during the ramp. While the effective temperatures extracted from the charge and spin susceptibilities and single-particle Green's function are relatively well defined in frequency space, they differ substantially from each other and depend on the momentum, which shows that a unique nonequilibrium temperature of a strongly perturbed weakly interacting system cannot be defined within a few hopping times after a ramp. How exactly the different effective temperatures approach the thermal value in the weak- and intermediate-correlation regime is an interesting subject for further studies, which however requires the implementation of memory-truncation techniques~\cite{PhysRevB.105.115146} to access the long-time dynamics. 


{\it Acknowledgments --}
We thank A.M-S Tremblay for useful conversations. The calculations have been performed on the Beo05 cluster at the University of Fribourg. This work was supported by ERC Consolidator Grant No.~724103.

\bibliography{Bibliography}

\clearpage

\begin{widetext}
\section*{Nonequilibrium Two-Particle Self-Consistent Approach --- Supplementary Materials}


In these supplementary notes, we provide benchmarks against equilibrium results available in the literature and some complementary nonequilibrium results, 
both for TPSC and TPSC+GG. Overall, TPSC and TPSC+GG agree well for temperatures larger than the renormalized classical crossover temperature $T_x$ in the weak-coupling regime $U\lesssim 3$.  However, the precursors of the antiferromagnetic (AFM) bands do not show up in the single-particle spectrum in TPSC+GG, in contrast to TPSC. 


\section{TPSC benchmarks}
In this section, we test our implementation by reproducing equilibrium results which have been published in the literature.

\begin{figure}[h!]
  \centering
    \includegraphics[width=0.5\linewidth]{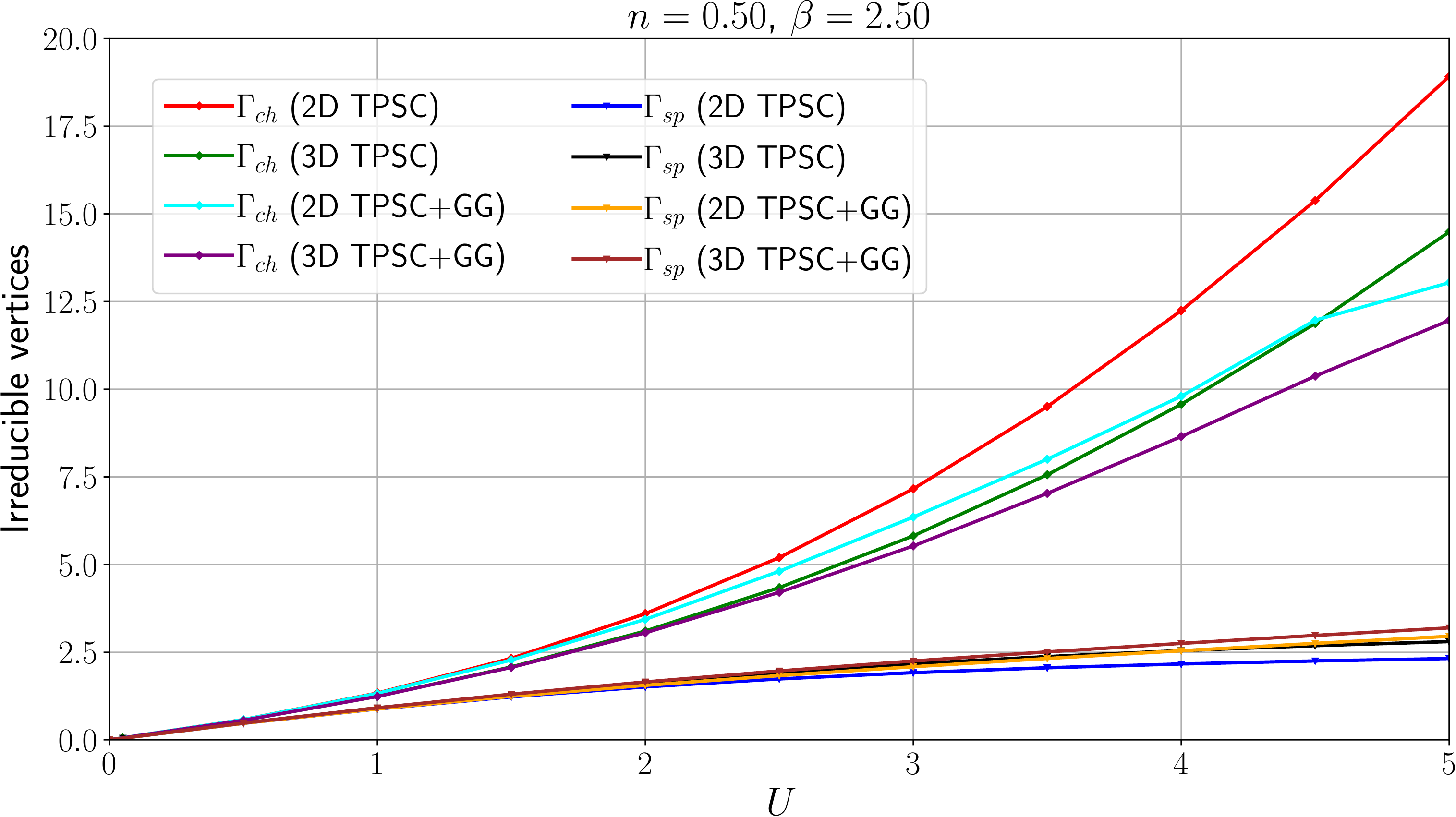}
      \caption{Spin and charge irreducible vertices as a function of bare interaction for the square (2D) and cubic (3D) lattices. The inverse temperature is $\beta=2.5$ and the spin-density per site is $n=0.5$. The TPSC+GG result for the charge vertex inflects at intermediate values of $U$, especially for the 2D case. This figure can be compared to Fig.~2 in Ref.~\cite{tpsc_1997}.}
  \label{fig:usp_uch_vs_u}
\end{figure} 

We first illustrate in Fig.~\ref{fig:usp_uch_vs_u} the equilibrium behavior of the spin and charge irreducible vertices as a function of the bare interaction for both the square and cubic lattices. Figure~\ref{fig:usp_uch_vs_u} shows that the spin irreducible vertex saturates at higher values of the bare interaction, due to the Kanamori-Brueckner screening~\cite{kanamori_brueckner_1963}. This screening implies that the crossover temperature into the renormalized classical regime $T_x$ saturates with increasing $U$~\cite{tpsc_1997}, as can be seen from Fig.~\ref{fig:static_spin_sus_fermi_surface}, which shows the spin-susceptibility of the two-dimensional model for increasing $U$ -- the up-turns become closer in temperature.

\begin{figure}[h!]
  \centering
    \includegraphics[width=0.5\linewidth]{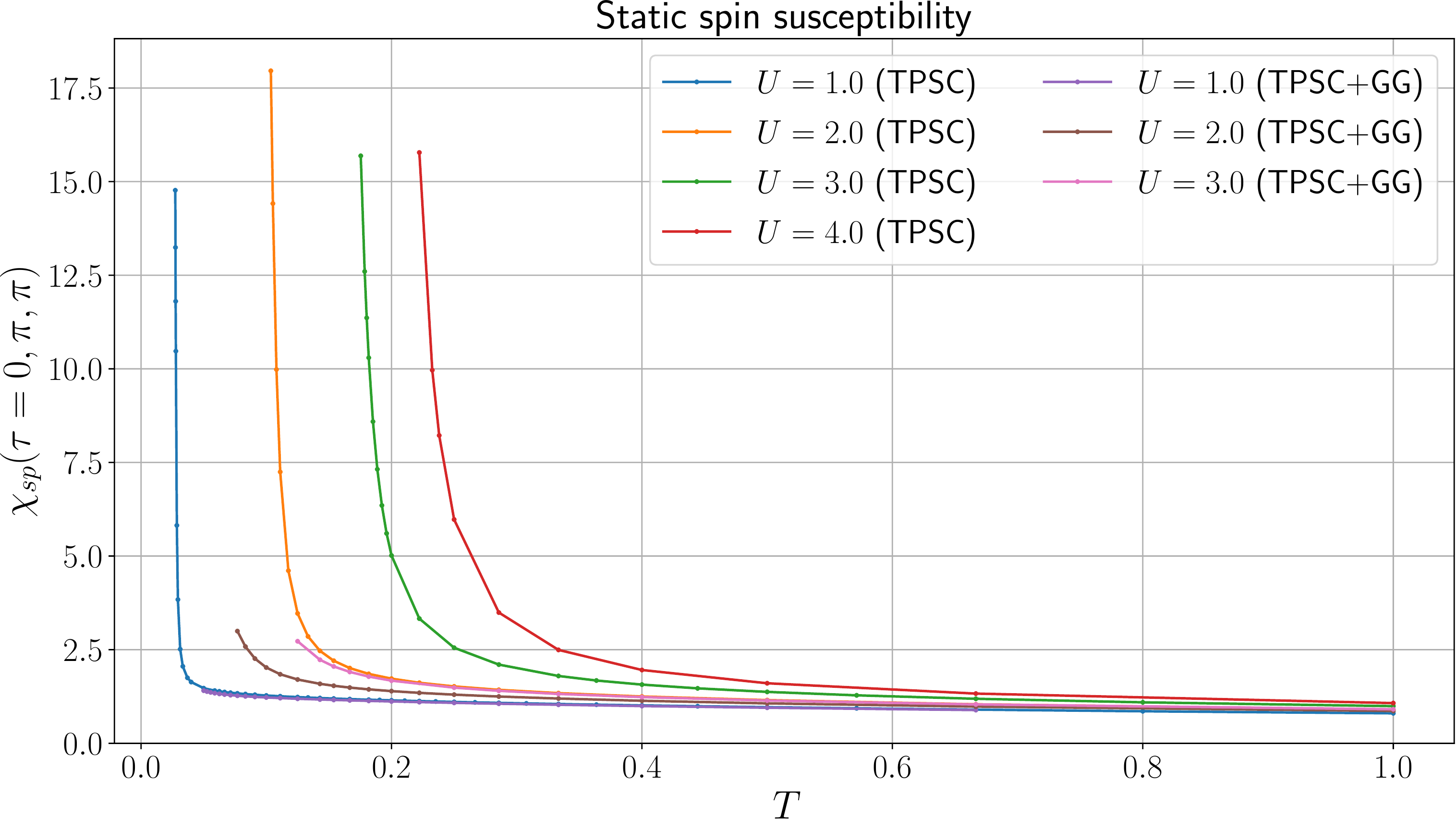}
      \caption{Static spin susceptibility of the 2D model at momentum $\mathbf{k}=(\pi,\pi)$ as a function of temperature for interactions $U=1,2,3$ and $4$. The filling per spin is $n=0.5$. The data points for TPSC+GG at $U=4$ are not shown since the solution becomes unstable at high temperature ($T\approx 0.3$). The red curve can be compared with Fig.~3 of Ref.~\cite{PhysRevB.49.13267_tpsc_1994}.
      }
  \label{fig:static_spin_sus_fermi_surface}
\end{figure}

The crossover temperature $T_x$ corresponds to the temperature where the static spin susceptibility starts shooting up, as illustrated in Fig.~\ref{fig:static_spin_sus_fermi_surface} for various values of the interaction using TPSC and TPSC+GG. Increasing the bare interaction increases the value of $T_x$ and it starts saturating at larger interaction values. As the bare interaction decreases, TPSC+GG drifts away from the TPSC results at lower temperatures. Also, the TPSC+GG results do not show a steep shooting-up of the static spin susceptibility at high temperature, like for TPSC, which is related to the observation that the precursor AFM bands do not show up in the single-particle spectra in the renormalized classical regime.

\begin{figure}[h!]
  \centering
    \includegraphics[width=0.5\linewidth]{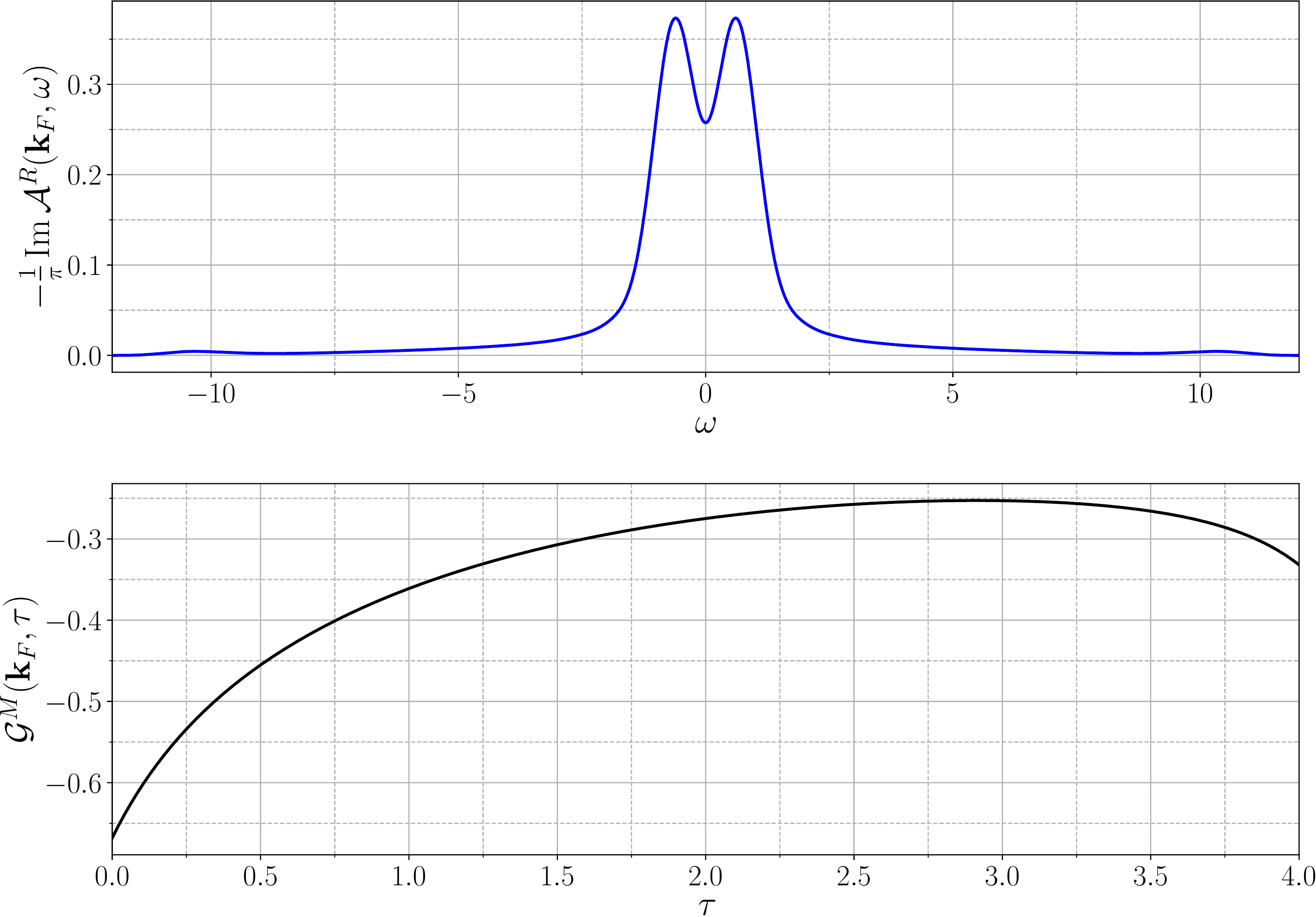}
      \caption{Top panel: TPSC electronic spectral density of the 2D model for spin-density $n=0.5$ and the Fermi surface momentum $\mathbf{k}_F=(0,\pi)$. The bare interaction is $U=4$ and the inverse temperature $\beta=5.88$. Bottom panel: TPSC Matsubara Green's function for $n=0.4375$ for the 2D model and the Fermi surface momentum $\mathbf{k}_F=(0,\pi)$. The bare interaction is $U=4$ and the inverse temperature $\beta=4$. These results can be compared with Fig.~9 in Ref.~\cite{tpsc_1997} (top panel) and the left panel of Fig.~1 in Ref.~\cite{Vilk_1996} (bottom panel). Since our calculations are implemented on the Kadanoff-Baym contour, spectral functions can be calculated directly by Fourier transformation, i.e. without analytical continuation.
      }
  \label{fig:spectral_components_fermi_surface}
\end{figure} 

In the renormalized classical regime, the growth of the spin fluctuations leads to a precursor of an AFM gap,
as can be seen in the top panel of Fig.~\ref{fig:spectral_components_fermi_surface}. The spin fluctuations destroy the Fermi-liquid quasiparticles above the zero-temperature phase transition in 2D (TPSC fulfils the Mermin-Wagner theorem)~\cite{tpsc_1997}. To observe the AFM pseudo-gap 
in the spectral function near the Fermi level, we need to dress the nonlocal Green's function with the TPSC self-energy [Eq.~(8) in the main text]. In the bottom panel of Fig.~\ref{fig:spectral_components_fermi_surface}, we show the Matsubara component of the dressed Green's function for the filling of $n=0.4375$ at the Fermi surface ($\mathbf{k}_F=(0,\pi)$). The results of Fig.~\ref{fig:spectral_components_fermi_surface} can be compared with Ref.~\cite{Vilk_1996} (left panel of Fig.~1) and Ref.~\cite{tpsc_1997} (Fig.~9).

\clearpage

\section{Equilibrium spectra}

As a reference for the difference spectra shown in the main text and in the following section, we plot in Fig.~\ref{fig:lesser_spin_and_charge_sus_equilibrium} the equilibrium TPSC+GG {\it lesser} spectra of the charge (top panels) and spin (bottom panels) susceptibilities for $U=1$ (left panels) and $U=3$ (right panels). At $U=3$, the difference between the charge and spin spectra is obvious: the charge susceptibility features an excitation gap away from $\mathbf{k}=(0,0)$, while the spin response is marked by predominant weight around $\omega\simeq 0$ in the vicinity of $\mathbf{k}=(\pi,\pi)$. At $U=1$, the charge and spin susceptibilities look more alike: by decreasing the Hubbard interaction, both the spin and charge response functions approach the Lindhard function for the square lattice.

\begin{figure}[h]
\begin{minipage}[h]{0.48\linewidth}
\begin{center}
\includegraphics[scale=0.8]{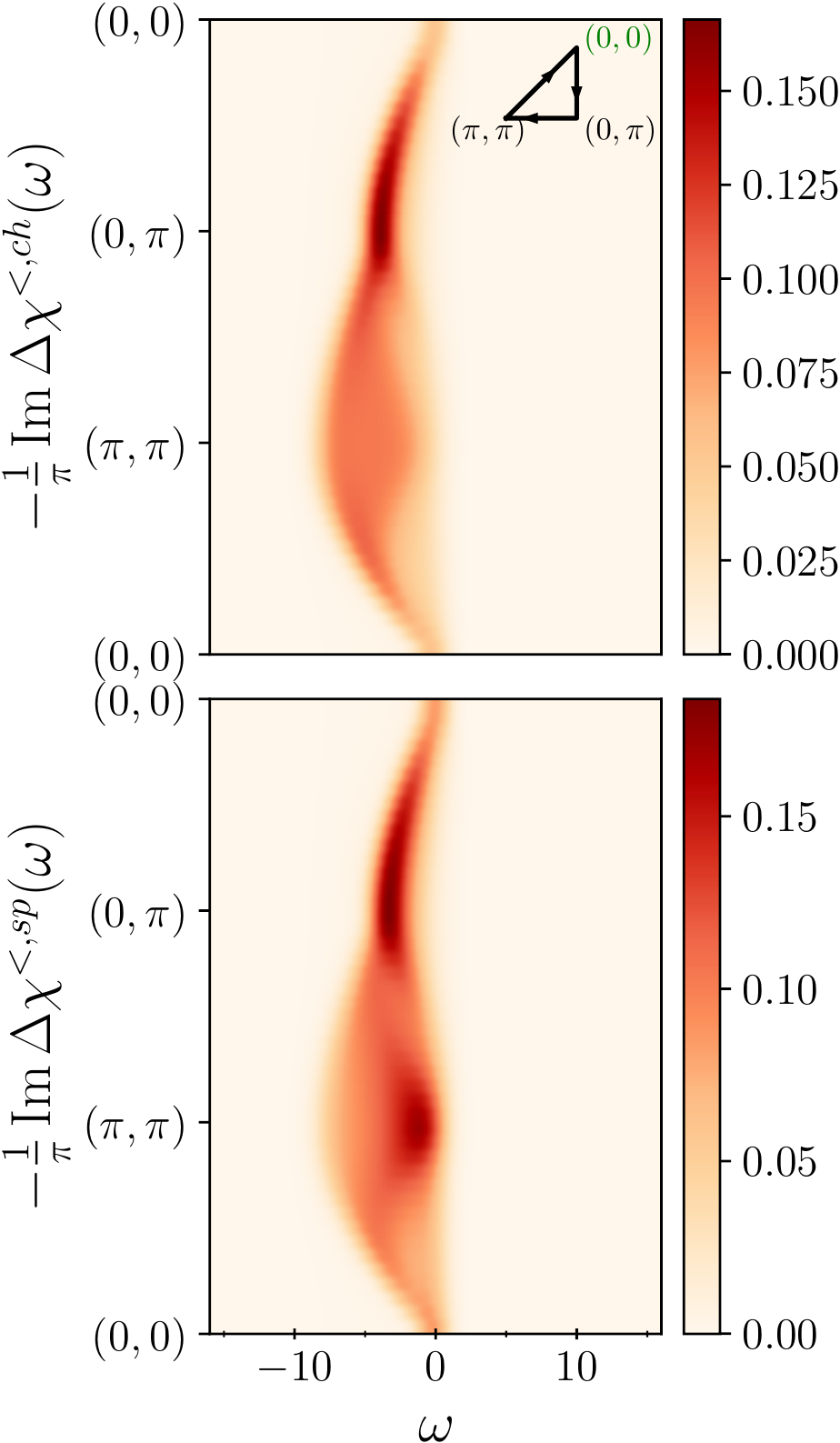} 
\end{center} 
\end{minipage}
\hfill
\vspace{0.1 cm}
\begin{minipage}[h]{0.48\linewidth}
\begin{center}
\includegraphics[scale=0.8]{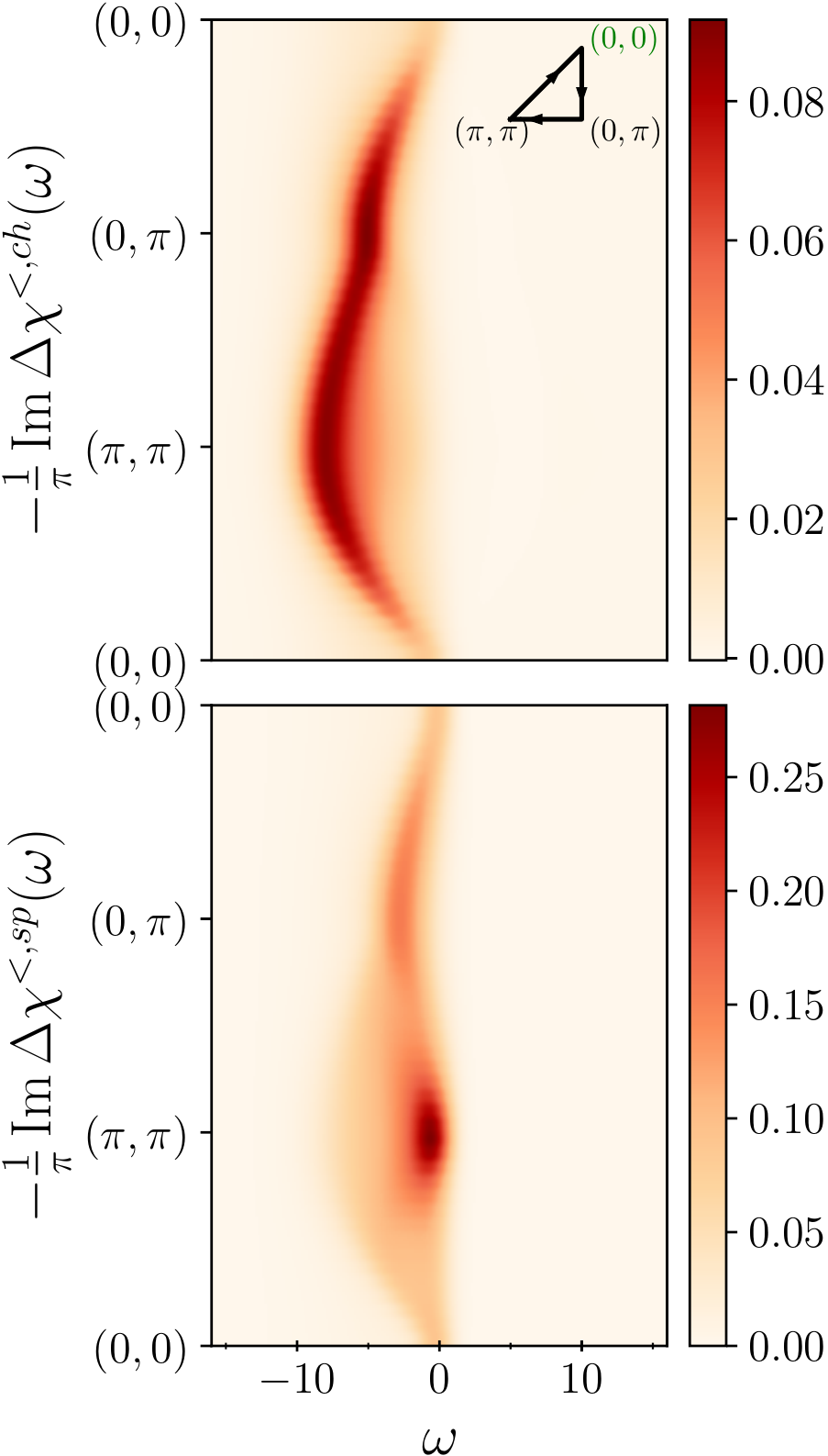} 
\end{center}
\end{minipage}
\caption{The imaginary parts of the \textit{lesser} component of the charge (top panels) and spin (bottom panels) susceptibilities, obtained by TPSC+GG. The left (right) panels show the equilibrium spectra for $U=1$ ($U=3$). The inverse temperature is $\beta=3$. The time window used for the Fourier transform is $\Delta t = 4$.}
\label{fig:lesser_spin_and_charge_sus_equilibrium}
\end{figure}

\clearpage

\section{Additional results for interaction ramps}
In this section, we show TPSC results which complement the TPSC+GG results presented in the main text, along with extra results obtained with TPSC+GG.

Figure~\ref{fig:lesser_spin_and_charge_sus} shows the evolution of the \textit{lesser} component of the spin susceptibility (left panels) and of the charge susceptibility (right panels) after an interaction ramp, obtained with TPSC+GG. The upper (lower) row of panels corresponds to an upward (downward) interaction ramp, as indicated in the insets. At $U=3$ and $\beta=3$, the system is in the vicinity of the AFM crossover, also called renormalized classical regime~\cite{tpsc_1997}. For that reason, as shown in the top left (bottom left) panel of Fig.~\ref{fig:lesser_spin_and_charge_sus}, the spin susceptibility at momentum $\mathbf{k}=(\pi,\pi)$ decreases (increases) strongly around $\omega\simeq 0$ when ramping from $U=3$ to $U=1$ ($U=1$ to $U=3$). At the same momentum, the \textit{lesser} component of the charge susceptibility exhibits a more qualitative change: as the bare interaction $U$ is decreased (bottom right panel), the sharp excitation at $\omega\simeq -8=-W$ spreads out to cover the energy region up to $\omega\approx 0$, reminiscent of the Lindhard function. During the ramp, a transient peak appears near $\omega\simeq 0$, indicating the availability of low-energy charge excitations. This peak also shows up as a weakly dispersing band in the top left panel of Fig.~\ref{fig:k_resolved_charge_vanilla} (red shades around $\omega\simeq 0$), which is for a faster ramp. Similar transient spectral features do not prominently show up during the up ramps, as can be seen in the top right panel.

\begin{figure}[h]
\begin{minipage}[h]{0.48\linewidth}
\begin{center}
\includegraphics[width=1\linewidth]{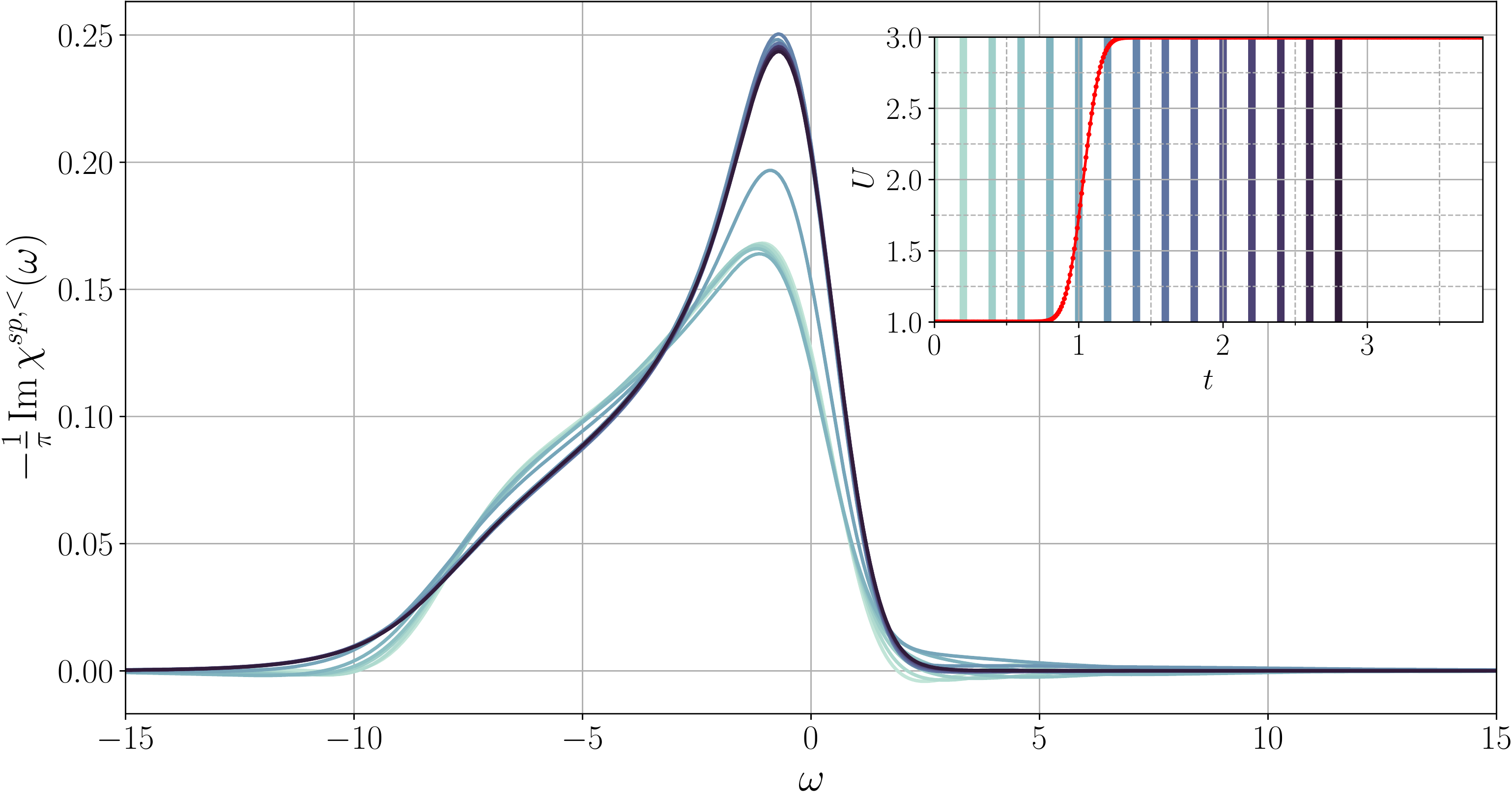} 
\end{center} 
\end{minipage}
\hfill
\vspace{0.1 cm}
\begin{minipage}[h]{0.48\linewidth}
\begin{center}
\includegraphics[width=1\linewidth]{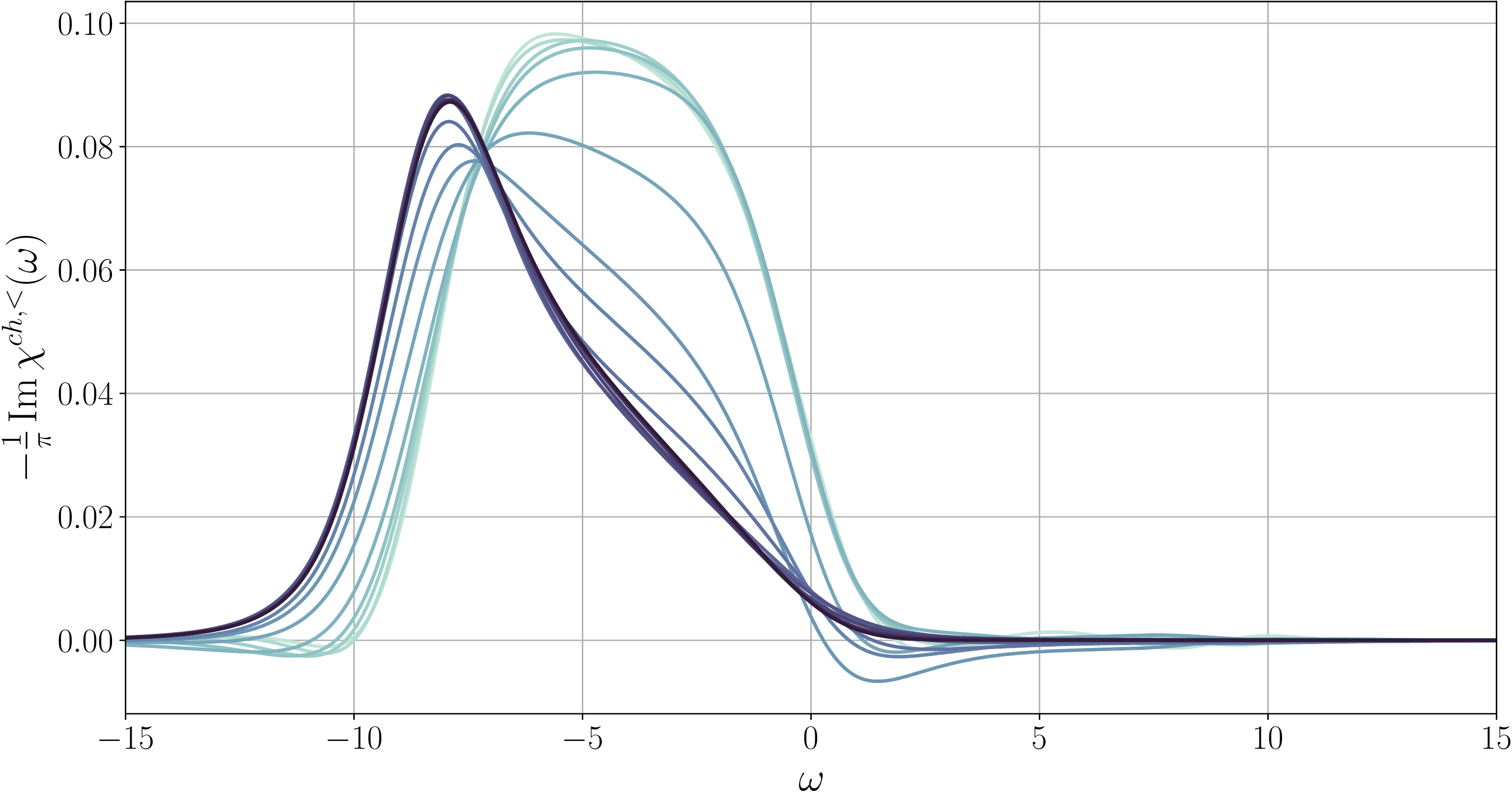} 
\end{center}
\end{minipage}
\vfill
\vspace{0.1 cm}
\begin{minipage}[h]{0.48\linewidth}
\begin{center}
\includegraphics[width=1\linewidth]{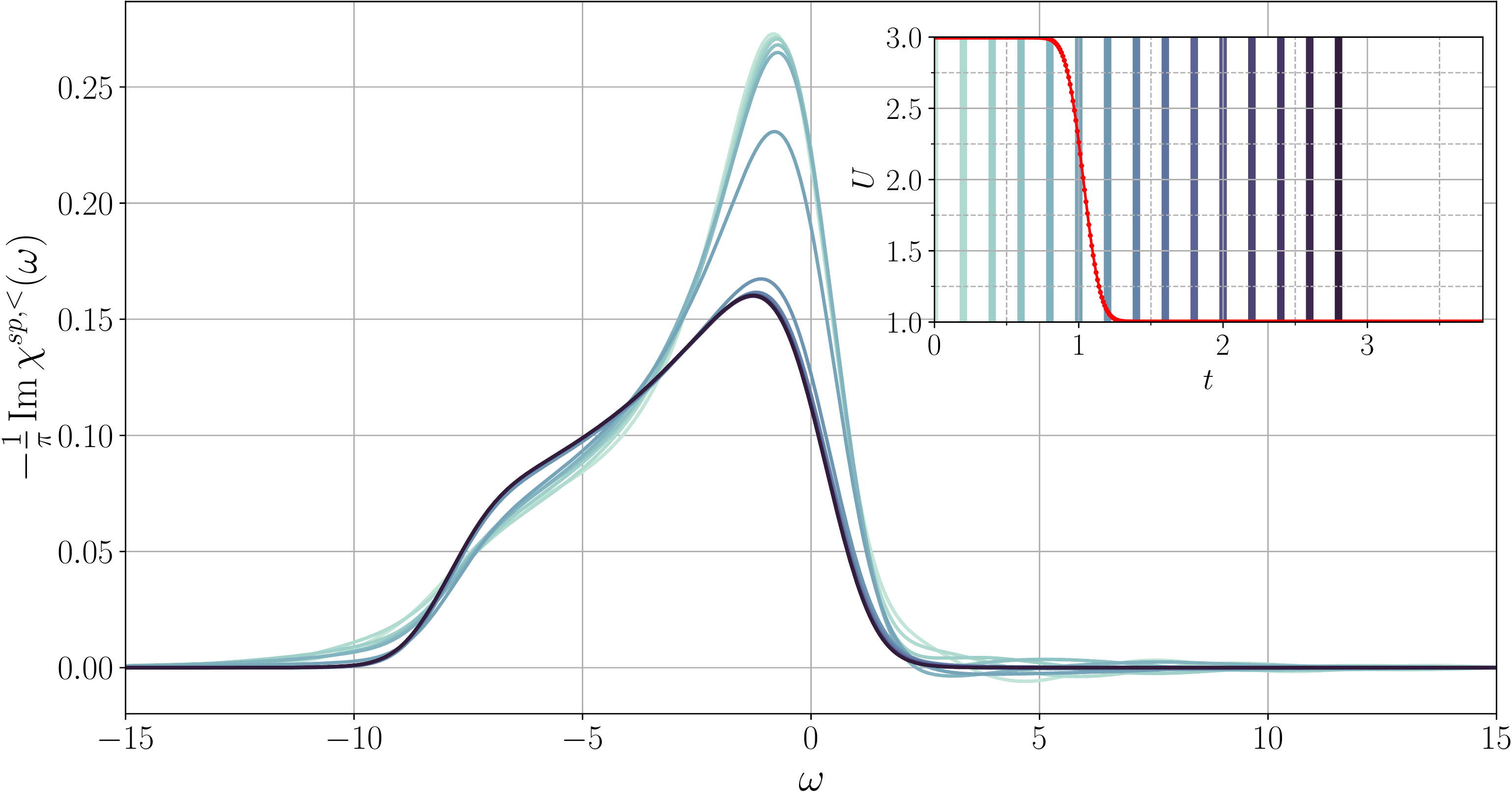} 
\end{center}
\end{minipage}
\hfill
\begin{minipage}[h]{0.48\linewidth}
\begin{center}
\includegraphics[width=1\linewidth]{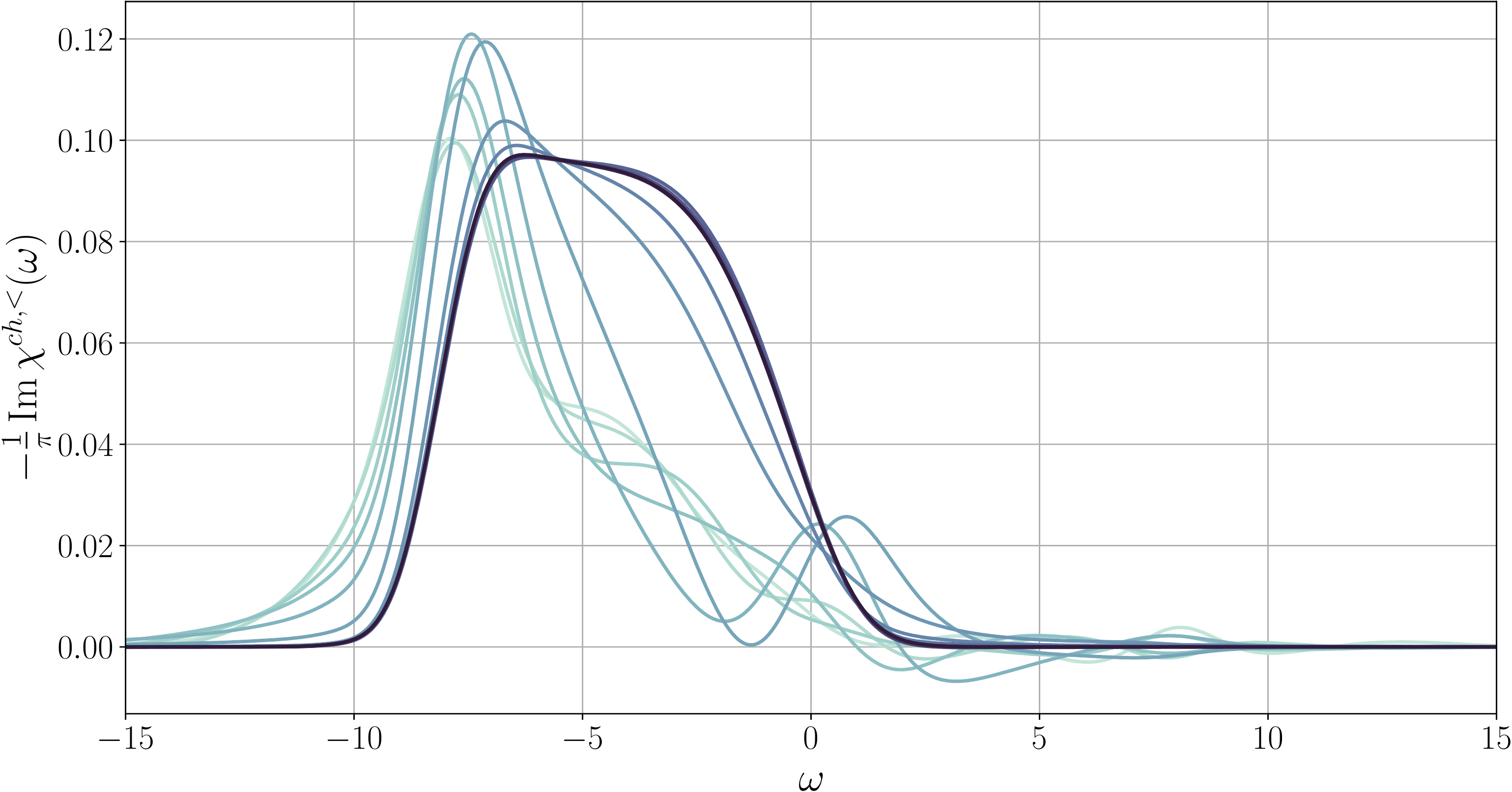} 
\end{center}
\end{minipage}
\caption{The imaginary parts of the \textit{lesser} component of the spin (left panels) and charge (right panels) susceptibilities (Eq.~(6) in the main text) for momentum $\mathbf{k}=(\pi,\pi)$. The top (bottom) panels display the evolution of the susceptibilities upon ramping the interaction up (down). The insets show the profiles of the interaction ramps with the vertical bars representing the times for which the spectra are calculated. The time window for the Fourier transformation is $\Delta t=4$.
}
\label{fig:lesser_spin_and_charge_sus}
\end{figure}

In Fig.~\ref{fig:k_resolved_charge_vanilla}, we plot the TPSC results for the difference spectra (\textit{lesser} components) of 
the charge (top panel) and spin (bottom panel) susceptibilities for the same interaction ramp as in Fig.~2 of the main text. The initial inverse temperature is also $\beta=3$. We notice that there is mainly a difference in the colorscale between the bottom panel of Fig.~\ref{fig:k_resolved_charge_vanilla} and the bottom panel of Fig.~2 in the main text (the latter showing TPSC+GG results). These differences arise due to the fact that TPSC overestimates the spin correlation growth. We obtain similar results with the two schemes for the charge susceptibility (top panels). 

\begin{figure}[h!]
  \centering
    \includegraphics[width=0.5\linewidth]{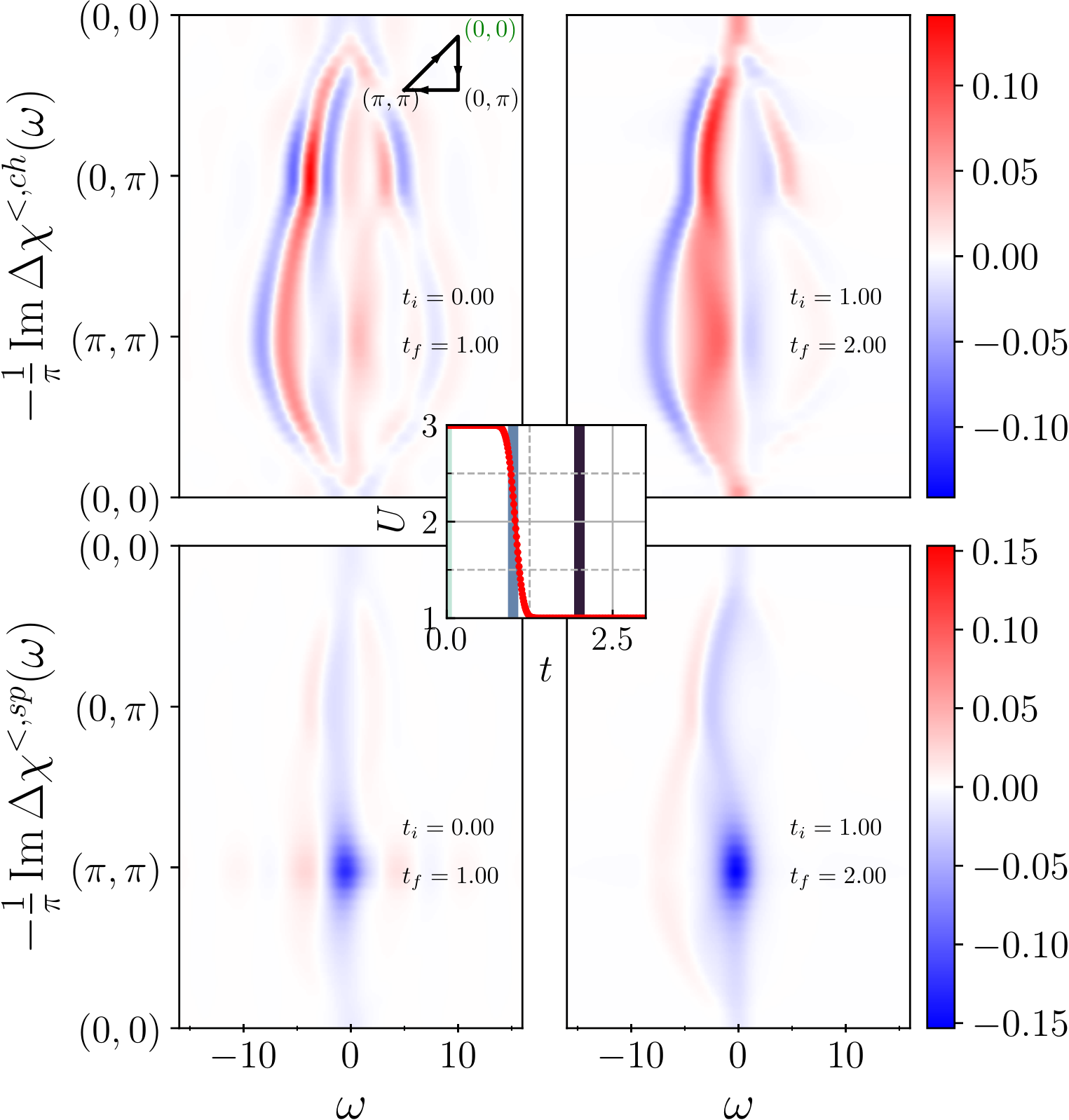}
      \caption{Difference spectra analogous to Fig.~2 in the main text, but calcuated with TPSC. The corresponding initial and final times $t_i$ and $t_f$ are indicated in each panel. Top (Bottom) panels: The spectral difference of the \textit{lesser} charge (spin) susceptibility after the interaction ramp shown in the inset (the vertical bars indicate the times $t_i$ and $t_f$). The inset black triangle illustrates the path in reciprocal space along which the spectra are displayed, with the green coordinates corresponding to the initial $\mathbf{k}$-point. The time window used for the Fourier transformation is $\Delta t=4$. Each row of panels uses the same colorscale.} 
\label{fig:k_resolved_charge_vanilla}
\end{figure}

In Fig.~\ref{fig:spectral_components_fermi_surface_vanilla}, the irreducible vertices, the double occupancy and the parameter $\alpha$ calculated within TPSC are illustrated for different ramp profiles and the initial inverse temperature $\beta=3$. The thermal inverse temperature for the slow (fast) ramp is $\beta_{\text{th}}=2.83$ ($\beta_{\text{th}}=2.71$). This is the equivalent of Fig.~1 in the main text (which is for TPSC+GG). As in Fig.~1 of the main text, we can see that the charge vertex approaches the thermal value later than the spin vertex and that the transient evolution of the charge vertex does not simply follow the ramp shape. In TPSC, after the ramp, oscillations appear in the charge vertex, but not in the spin vertex. Those oscillations have a frequency of $\omega_\text{osc}\simeq\frac{2\pi}{W}$, where $W=8$ corresponds to the bandwidth of the square lattice. We verified that these oscillations remain present when changing the parameters. The thermal values expected for the different quantities after the slow and fast ramps are indicated by the arrows (the arrows essentially overlap for the two ramp profiles). $\Gamma_{\text{ch}}$ and $\Gamma_{\text{sp}}$ thermalize at values which are farther apart from each other as compared to the TPSC results (Fig.~1 in the main text). As in the main text, these reference data are calculated from the total energy after the ramp, and differ very little between the two ramps. 

\begin{figure}[h!]
  \centering
    \includegraphics[width=0.5\linewidth]{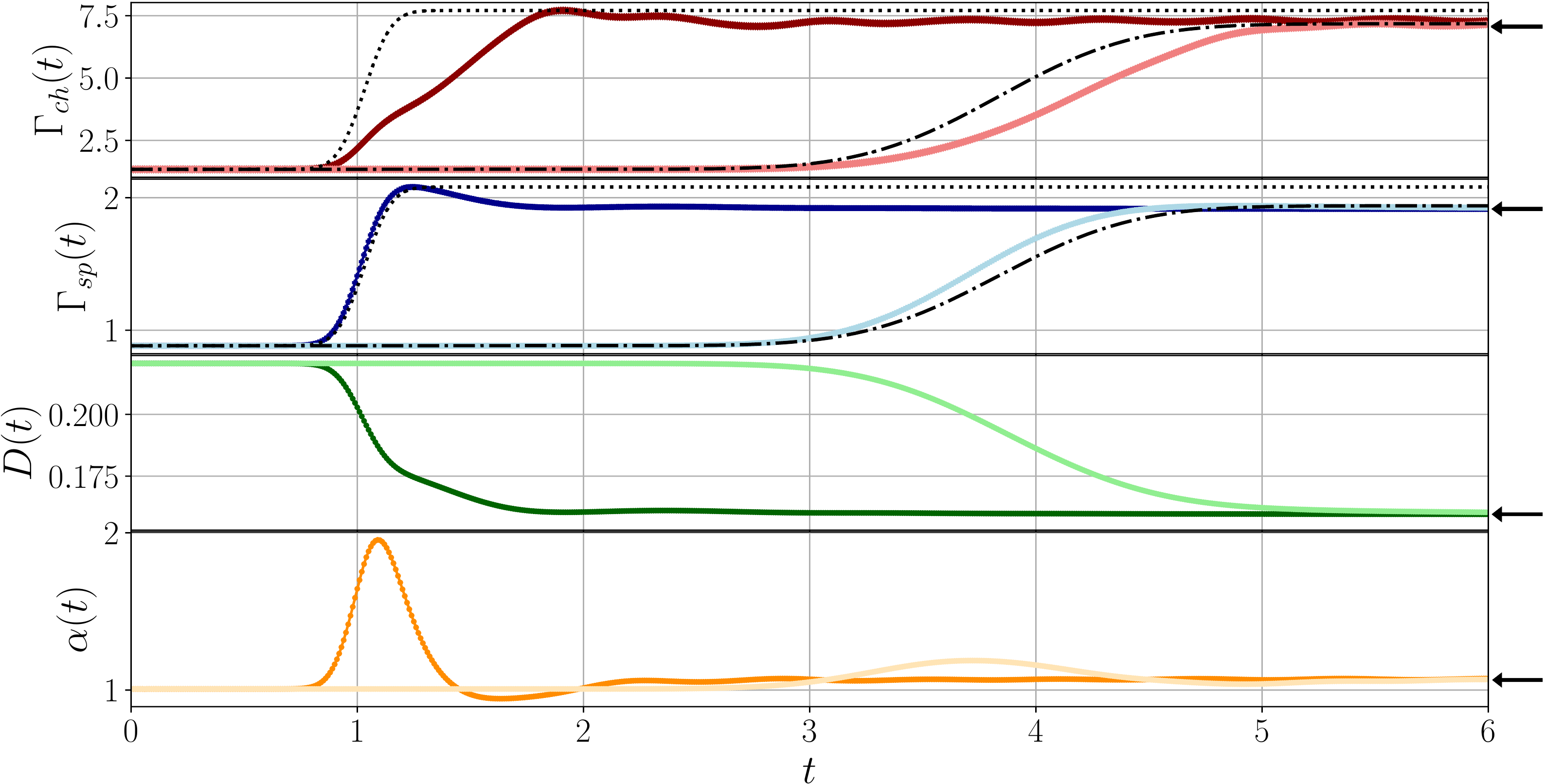}
      \caption{Results analogous to Fig.~1 in the maint text, but calculated with TPSC. The quantities plotted in light colors are associated with the smooth interaction ramp and those in dark colors with the sharp interaction ramp. The dotted lines represent the interaction ramps in arbitrary units. Top panel: charge irreducible vertex $\Gamma^\text{ch}$. Second panel: spin irreducible vertex $\Gamma^\text{sp}$. Third panel: Double occupancy $D(t)$. Bottom panel: $\alpha$ parameter enforcing the sum-rule Eq.~(9) in the main text. The thermal values for the two different ramps are almost indistinguishable.}
  \label{fig:spectral_components_fermi_surface_vanilla}
\end{figure}

Finally, in Fig.~\ref{fig:up_quench_time_diff_lesser_susceptibilities}, we show in the left subplot the time differences of the \textit{lesser} spectral weight of the spin (left panels) and charge (right panels) susceptibilities calculated within TPSC+GG for an up ramp from $U=1$ to $U=3$. The up ramp is shown in the inset panel and is the inverse of the down ramp considered in Fig.~2 of the main text. 

In the right subplot of Fig.~\ref{fig:up_quench_time_diff_lesser_susceptibilities} we show TPSC results for the spin and charge susceptibilities for the same up ramp. A clear observation from Fig.~\ref{fig:up_quench_time_diff_lesser_susceptibilities} is that the changes in the spectra are substantially larger within TPSC than within TPSC+GG; even more so than for the down ramp (see Fig.~2 of the main text and Fig.~\ref{fig:k_resolved_charge_vanilla}). Also, the time differences for the first and second half of the ramp are more similar in TPSC+GG than in TPSC. Furthermore, it seems that the larger heating effect in TPSC+GG produces smaller time difference spectra for the up ramp. When comparing the difference spectra for the down and up ramps, one notices that the transient charge excitations near $\omega\simeq 0$ mainly show up during the down ramp. In both schemes, the up ramp brings the system close to the renormalized classical regime.

\begin{figure}[h]
\begin{minipage}[h]{0.48\linewidth}
\begin{center}
\includegraphics[width=1\linewidth]{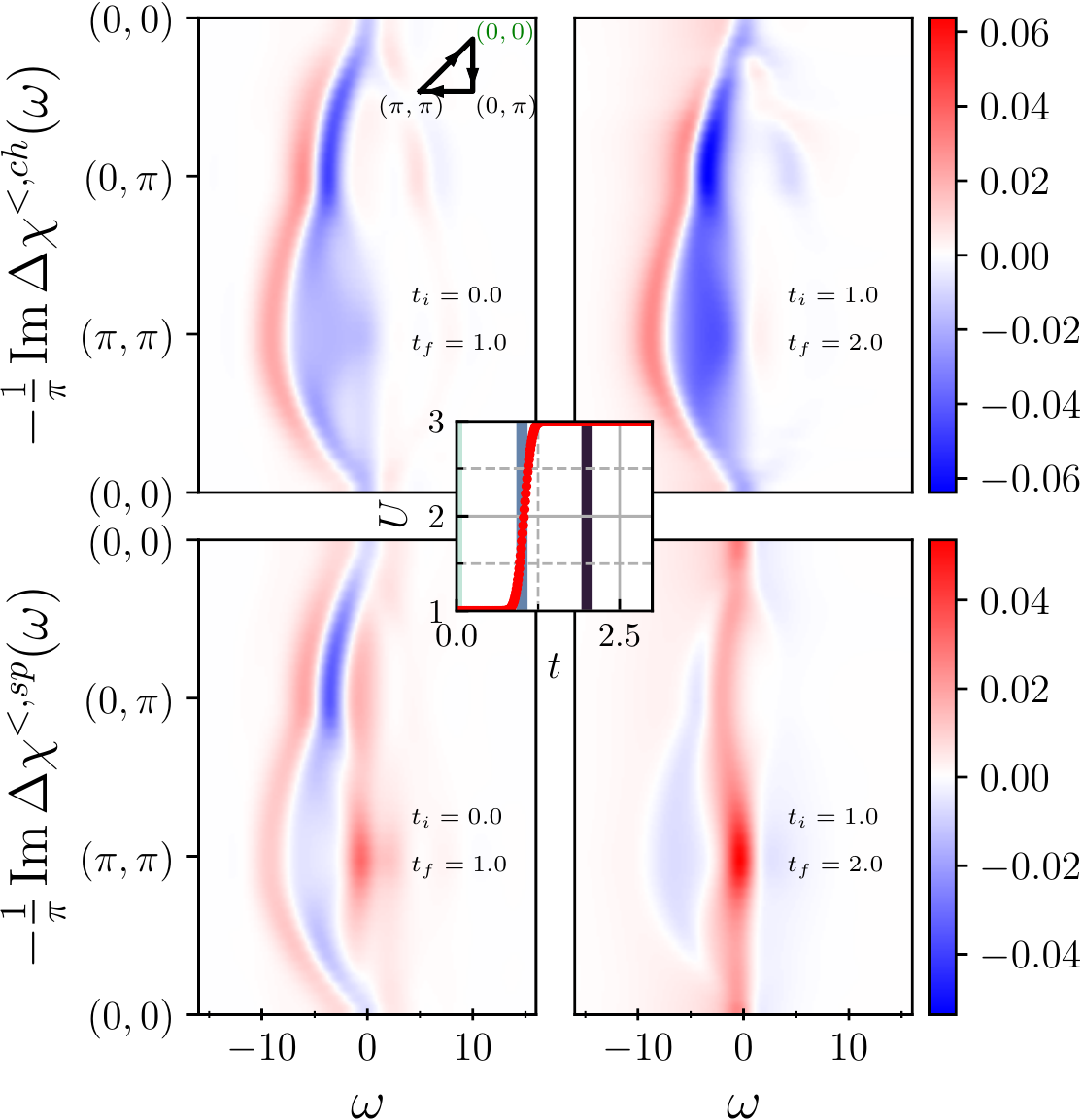} 
\end{center} 
\end{minipage}
\hfill
\vspace{0.1 cm}
\begin{minipage}[h]{0.48\linewidth}
\begin{center}
\includegraphics[width=1\linewidth]{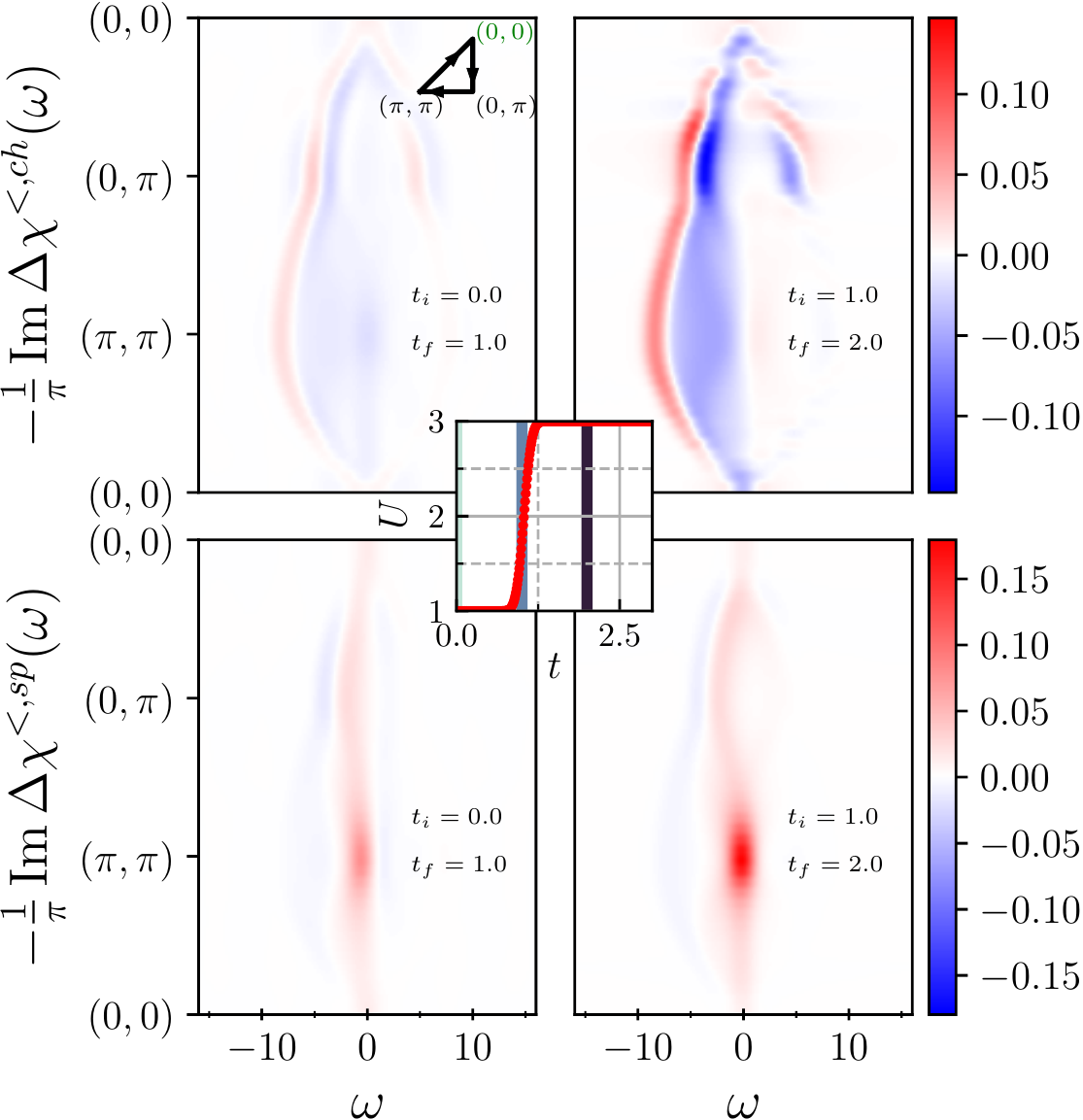} 
\end{center}
\end{minipage}
\caption{Top (Bottom) panels: Difference spectra of the lesser component of the charge (spin) susceptibility after the interaction ramp from $U=1$ to $U=3$ shown in the inset. The left (right) subplot shows the results obtained using TPSC+GG (TPSC). The time window employed in the Fourier transformation is $\Delta t=4$. Each row of panels uses the same colorscale.}
\label{fig:up_quench_time_diff_lesser_susceptibilities}
\end{figure}
\end{widetext}

\end{document}